\documentclass[tc, manuscript]{copernicus} 

\begin{document}

\nolinenumbers

\title{Wave-induced erosion and notch development at the Fimbul Ice Shelf front}


\Author[1][wenjun.lu@ntnu.no]{Wenjun}{Lu} 
\Author[3]{Lotte}{Wendt}
\Author[2]{Behnam}{Ghadimi}
\Author[1]{Shovon}{Jubair}
\Author[2]{Dominique}{Mouaze}
\Author[2]{Marianne}{Font}
\Author[2]{R\'emi}{Lambert}
\Author[3]{Harvey}{Goodwin}
\Author[3]{Geir}{Moholdt}
\Author[1]{Raed}{Lubbad}
\Author[1]{Sveinung}{L\o{}set}

\affil[1]{Department of Civil and Environmental Engineering, Faculty of Engineering,
Norwegian University of Science and Technology (NTNU), Trondheim, Norway}
\affil[2]{M2C, UMR CNRS 6143, Universit\'e de Caen Normandie, Caen, France}
\affil[3]{Norwegian Polar Institute, Fram Centre, Troms\o{}, Norway}




\runningtitle{Wave-induced ice-shelf erosion: spectral and breaking-wave extensions}

\runningauthor{Lu et al.}

\received{}
\pubdiscuss{} 
\revised{}
\accepted{}
\published{}


\firstpage{1}

\maketitle

\begin{abstract}
Ocean waves erode the waterline of Antarctic ice-shelf fronts and carve a thermo-erosional notch. The notch is hidden from satellites, yet it preconditions front collapse and footloose calving, so the link between notch growth and observable front retreat matters for how wave-exposed ice shelves lose mass. We study this link at the Fimbul Ice Shelf, East Antarctica, in January--February 2024. The widely used \citet{white1980} erosion formulation is extended in two directions: a component-wise spectral treatment of irregular seas, and a breaking-aware wave profile that accounts for shoaling over the submerged ice foot revealed by remotely operated vehicle (ROV) profiling. Forced with hourly ERA5 waves, the extended model predicts about 110~m of cumulative waterline erosion over the matched 7~January--27~February window. Satellite observations show considerably more retreat: about 264~m from an S1-guided Sentinel-2 plateau-break method, and 266~m from manually digitised Sentinel-1 fronts over a slightly longer window. Under the baseline assumptions ($\alpha=1$, $\Delta T_{\mathrm{wi}}=1$~$^\circ$C), the model therefore falls short of the observed retreat by a factor of about 2.4. Part of this residual may be hydrodynamic, since post-breaking turbulence is not represented; part may be mechanical, because collapse and footloose calving can convert notch erosion into larger observable retreat. Unmeasured near-ice thermal driving remains a first-order uncertainty, and front-position observations alone cannot separate these contributions.
\end{abstract}



\section{Introduction}

\begin{figure}[t]
\centering
\includegraphics[height=0.62\textheight,keepaspectratio]{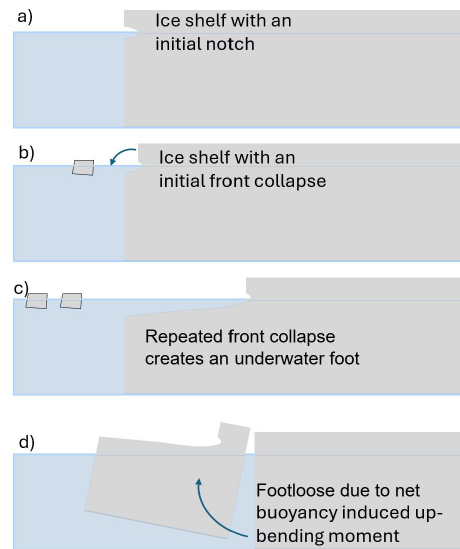}
\caption{Schematic sequence linking wave erosion to two modes of calving response.
(a)~Wave-driven heat transfer creates a thermo-erosional notch at the waterline.
(b)~Deepening of the notch removes support from the overhanging slab and can trigger
front collapse. (c)~Repeated collapse and retreat of the above-water face leave a
submerged ice foot extending seaward. (d)~The net buoyancy of this foot applies a bending
moment to the ice shelf and may promote footloose detachment. The geometry is schematic
and not to scale. This study quantifies the notch-erosion forcing represented in panel~(a).}
\label{fig:erosion_schematic}
\end{figure}

Ice shelf calving accounts for approximately half of the mass loss from the Antarctic Ice Sheet \citep{sartore2025}. While giant tabular calving events receive significant attention, small-scale ``front collapse'' events (termed \emph{spalling} in the glaciological literature; \citealp{scambos2005, scambos2008})---on the sub-kilometre scale---are increasingly recognised as a continuous driver of ice-front retreat \citep{sartore2025}. Wave-induced erosion at the waterline can contribute to these collapses by creating a thermo-erosional notch. Notch growth alters the stress distribution within the overlying ice slab and may promote calving once a critical geometry is reached \citep{white1980}. Furthermore, repeated front collapse leaves behind a submerged ice foot, whose uncompensated buoyancy exerts a bending moment that can promote large-scale ``footloose'' calving \citep{wagner2014footloose,sartore2025}. Both failure mechanisms are illustrated in Fig.~\ref{fig:erosion_schematic}.

Despite the fundamental role of notch development, estimating its rate under ocean wave forcing remains a major physical challenge. The physics are governed by advective heat transfer, which strongly depends on the hydrodynamic regime. Oscillatory wave motion is thought to be the dominant heat delivery mechanism at wave-exposed ice fronts, by analogy to iceberg deterioration \citep{RN1481}, and wave-erosion formulations of this type have recently been applied at the Ross Ice Shelf front \citep{sartore2025}. At Fimbulisen, solar-heated surface water reaches the ice front in summer \citep{zhou2014asw}, supplying the thermal forcing on which wave-driven transfer can act.

\begin{figure}[t]
\includegraphics[width=\textwidth]{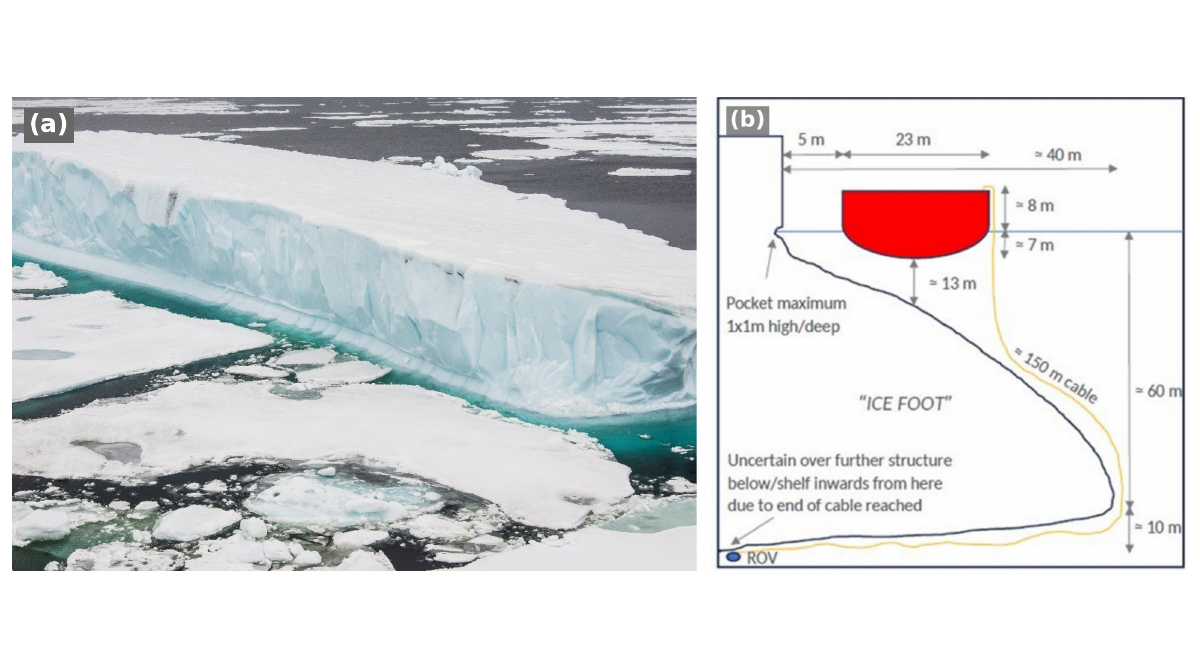}
\caption{(a)~An iceberg in the Greenland Sea showing pronounced waterline necking during
the Oden Arctic Research Cruise (OATRC) 2013. For a freely drifting iceberg, wave erosion
creates a simple waterline notch with no extended submarine foot, conditions well-suited
to the original \citet{white1980} formulation. (b)~Underwater profile of a submerged
ice foot at the Fimbulisen calving face profiled by a Remotely Operated Vehicle (ROV) in
2024. Unlike the simple waterline neck in panel~(a), the observed profile forms an
elongated submerged ramp beneath the frontal region. Incoming waves can shoal and break
as they propagate over this geometry, providing the physical motivation for the
breaking-aware model extension developed in this study.}
\label{fig:notch_photo}
\end{figure}

Currently, the operational standard for quantifying wave-driven erosion on a frozen body is the empirical linear wave theory (LWT) approach, often termed the ``White-lineage'' \citep{white1980}, which uses an empirical Stanton-number closure to parameterise heat transfer. While widely used for iceberg decay models \citep{RN1481}, it has significant uncertainties when applied to Antarctic ice shelves. It assumes regular monochromatic waves and ignores wave shoaling and breaking that become non-negligible at ice-shelf fronts where the submerged ice foot creates a shallow-water ramp (Fig.~\ref{fig:notch_photo}b). In addition, the standard practice of substituting bulk parameters ($H_s$ and $T_p$) directly into White's formula does not preserve the component-wise response of an irregular sea \citep{RN1845}. Although such spectral extensions of turbulent wave-current boundary-layer theory have been developed for friction quantification, wave-induced melting is still commonly evaluated from bulk sea-state parameters. Both limitations are discussed in Sect.~\ref{sec:gaps} and addressed in Sect.~\ref{sec:methods}. Meanwhile, existing ice-shelf parameterisations, such as the wind-driven iceberg-decay approach of \citet{gladstone2001}, have been suggested to overestimate erosion by an order of magnitude when applied to the Ross Ice Shelf front \citep{sartore2025}.

In this paper, we focus on extending and evaluating the empirical LWT framework at the wave-exposed Fimbul Ice Shelf (FIS) during the austral summer of January--February 2024. We extend the \citet{white1980} formulation in two critical ways: first, through a component-wise treatment of irregular seas; and second, by introducing a breaking-aware velocity profile that accounts for shoaling amplification over the submerged ice foot. This second extension is motivated by in situ remotely operated vehicle (ROV) profiling of the ice-shelf base, which reveals an elongated submarine foot extending seaward of the calving face, creating a shallow-water ramp where incoming waves can shoal and potentially break before impacting the ice face (Fig.~\ref{fig:notch_photo}b).

By comparing our extended LWT empirical model with observed retreat from Sentinel-1 SAR and an S1-guided Sentinel-2 plateau-break algorithm, we test the limits of current wave-erosion formulations. Under the baseline thermal and calibration assumptions, the modelled waterline erosion is smaller than the observed front retreat. The residual may reflect coupled post-breaking surf-zone hydrodynamics and fracture-controlled front collapse, including footloose amplification, while uncertainty in the unmeasured near-ice thermal driving limits quantitative attribution.

\section{Theoretical background: the modified wave-induced melting formulation}
\label{sec:background}

The theoretical foundation of this study is the modified White (1980) formulation derived, calibrated, and evaluated in the companion paper \citet{lu2025} (hereafter Paper~1). Paper~1 derives a depth-resolved erosion model for a vertical ice face under wave forcing and provides a laboratory calibration of the key empirical coefficient against experiments at the Caen M2C CNRS wave tank (Caen, France). We summarise the governing equations to the extent needed for the spectral and breaking-wave extensions introduced in Sect.~\ref{sec:methods}.

\subsection{Governing equations}
\label{sec:governing}

Wave-induced melting at a vertical ice face is treated as an advective heat-transfer problem in which oscillatory wave motion delivers heat to the ice surface. Under the assumption that all heat delivered is instantaneously expended in melting, the depth-resolved erosion rate derived and calibrated in Paper~1 is

\begin{equation}
\frac{T}{H}V_m
=
\alpha\cdot 0.0027\,\cos\phi
\left(\frac{\cosh[k(z+d)]}{\sinh(kd)}\right)^{0.8}
\frac{\left(\dfrac{k_s}{H}\right)^{0.2}}
{\mathrm{Re}_a(z)^{0.1}
+
\dfrac{3.84}{\sqrt{2}}\left(\mathrm{Pr}^{0.68}-1\right)}
\;\Delta T_{\mathrm{wi}} ,
\label{eq:Vm_full}
\end{equation}

where $k_s$ (m) is the ice-surface roughness length, $\mathrm{Pr}$ (dimensionless) is the Prandtl number of water, $\Delta T_{\mathrm{wi}}$ (K) is the water--ice thermal driving, and $\alpha$ (dimensionless) is a calibration coefficient (see below). The quantity $\mathrm{Re}_a(z) = u_m(z)\,a_{1m}(z)/\nu$ is the depth-dependent amplitude Reynolds number (dimensionless), which requires the horizontal orbital velocity amplitude

\begin{equation}
u_m(z) = \frac{\pi H}{T}\cos\phi\,\frac{\cosh[k(z+d)]}{\sinh(kd)},
\label{eq:um}
\end{equation}

and the orbital excursion amplitude

\begin{equation}
a_{1m}(z) = \frac{H}{2}\,\frac{\cosh[k(z+d)]}{\sinh(kd)},
\label{eq:a1m}
\end{equation}

where $H$ (m) is wave height, $T$ (s) is wave period, $k$ (m$^{-1}$) is wavenumber (from the dispersion relation $\omega^2 = gk\tanh(kd)$), $\phi$ (rad) is the angle of wave incidence, $d$ (m) is water depth, and $\nu$ (m$^2$\,s$^{-1}$) is the kinematic viscosity of water. The vertical coordinate $z$ is defined in the coordinate system introduced in Sect.~\ref{sec:methods}. Representative values of all physical constants are listed in Table~\ref{tab:constants}.

The depth-resolved field-scale equation (Eq.~\eqref{eq:Vm_full}) uses the compact power-law approximation to the Jonsson rough-wall friction coefficient. Paper~1 shows that this approximation reproduces the full Lambert-$W$ solution to within approximately 7\% over $a_{1m}/k_s \approx 30$--$300$, the field-scale excursion-to-roughness regime considered here. The full Jonsson/Nunner closure is retained in the independent irregular-wave equivalence test in Appendix~\ref{app:equivalence}. The depth-dependent velocity profile function in Eq.~\eqref{eq:Vm_full} is

\begin{equation}
R(z;\,k,\,d) = \frac{\cosh[k(z+d)]}{\sinh(kd)},
\label{eq:Rz_gen}
\end{equation}

with the corresponding amplitude Reynolds number

\begin{equation}
\mathrm{Re}_a(z;\,k,\,d) = \frac{\pi H^2}{2T\nu}\cdot R(z;\,k,\,d)^2.
\label{eq:Rea_gen}
\end{equation}

In deep water, shoaling and breaking are absent and the water depth is effectively infinite. The depth factors then reduce to the deep-water limit,

\begin{equation}
R_0(z) = e^{k_0 z},
\qquad
k_0 = \frac{\omega^2}{g} = \frac{4\pi^2}{gT^2}.
\label{eq:R0_dw}
\end{equation}

Equation~\eqref{eq:Vm_full} with $R(z;\,k,\,d)$ replaced by $R_0(z)$ and the corresponding simplified $\mathrm{Re}_a$ is the governing form under the deep-water simplification. 

When wave shoaling and breaking are considered, the full forms $R(z;\,k,\,d)$ and $\mathrm{Re}_a(z;\,k,\,d)$ are retained, with $k$ and $d$ varying as the wave propagates into shallower water. 

The calibration coefficient $\alpha$ in Eq.~\eqref{eq:Vm_full} was determined from the Caen wave-flume experiments in Paper~1. The two valid test conditions yield $\alpha = 0.684$ (Test~1) and $\alpha = 0.612$ (Test~2), giving a representative working value of $\alpha \approx 0.648$. Under the deep-water approximation, this calibrated value yields a baseline coefficient $C_{\mathrm{nb}} \approx 1.35\times10^{-4}$~K$^{-1}$, which is in close agreement with the empirical value of $1.46\times10^{-4}$~K$^{-1}$ recommended by \citet{white1980}. In this study we set $\alpha = 1$, giving $C_{\mathrm{nb}} = 2.09\times10^{-4}$~K$^{-1}$. At fixed thermal driving, this uncalibrated baseline produces more erosion and a smaller predictive shortfall than the calibrated value. The sensitivity is discussed in Sect.~\ref{sec:turbulence_gap}.

\subsection{Identified gaps for field application}
\label{sec:gaps}

Two fundamental simplifications in Eq.~\eqref{eq:Vm_full} limit its direct applicability to field conditions at an ice-shelf front, and motivate the model extensions in Sect.~\ref{sec:methods}.

First, the formulation assumes a single regular monochromatic wave characterised by one height $H$ and one period $T$. Real ocean wave fields are irregular and stochastic: energy is distributed across a range of frequencies and wave heights, and the nonlinear dependence of the erosion rate on wave height means that substituting bulk parameters such as $H_s$ and $T_p$ directly into Eq.~\eqref{eq:Vm_full} does not preserve the wave-population average. A component-wise spectral treatment is therefore required.

Second, the formulation ignores wave shoaling and breaking. For a freely drifting iceberg in deep water, this simplification is defensible, as the surrounding water depth always remains large relative to the wave amplitude. At an ice-shelf front, however, repeated front-collapse events leave behind a submerged ice foot that creates a shallow-water ramp. Incoming waves shoal over this ramp and may break before reaching the ice face, amplifying the orbital velocities and the associated thermal forcing. A breaking-aware model extension is therefore required to represent field conditions at a wave-exposed ice shelf.

\begin{table}[t]
\caption{Representative physical constants used in the modified White formulation (Eq.~\eqref{eq:Vm_full}).}
\label{tab:constants}
\centering
\begin{tabular}{lllll}
\hline
Symbol & Quantity & Value used & Units & Note \\
\hline
$\rho_w$ & Water density & 1025 & kg\,m$^{-3}$ & representative seawater \\
$\rho_i$ & Ice density & 917 & kg\,m$^{-3}$ & glacier ice \\
$c_p$ & Specific heat of water & 3985 & J\,kg$^{-1}$\,K$^{-1}$ & seawater \\
$\Gamma$ & Latent heat of fusion & $3.34\times10^{5}$ & J\,kg$^{-1}$ & ice \\
$\nu$ & Kinematic viscosity & $1.3\times10^{-6}$ & m$^{2}$\,s$^{-1}$ & value used in the model \\
$\mathrm{Pr}$ & Prandtl number & 13 & -- & seawater (order) \\
\hline
\end{tabular}
\end{table}

The calculation holds these properties fixed. Temperature- and salinity-dependent variations in density and viscosity, together with stratification produced by near-ice meltwater, are not resolved at the present model order. Their influence should be assessed with near-ice temperature and salinity profiles in future field measurements.

\section{Methodology}
\label{sec:methods}

\begin{figure}[t]
\centering
\includegraphics[width=\textwidth]{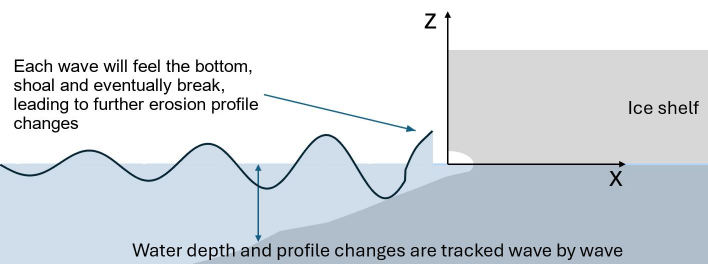}
\caption{Coordinate system and model geometry. The vertical coordinate $z$ (m) is measured
upward from the mean waterline ($z = 0$, positive upward). As the thermo-erosional notch
develops, incoming waves propagate over the ice-foot ramp, shoal, and may break at depth
$d_b$. The breaking-aware model computes the shoaling-amplified orbital velocities within
the shoaling zone $z \in [0, -d_b]$ and retains the non-breaking profile below.}
\label{fig:coordinates}
\end{figure}

Sect.~\ref{sec:gaps} identified two gaps. We close them in turn: Sect.~\ref{sec:irregular} extends the model from regular to irregular seas, and Sect.~\ref{sec:breaking} from non-breaking to breaking-aware wave profiles. We then describe the field site, forcing data, and satellite-based observations against which the models are evaluated. All theoretical development uses the coordinate system shown in Fig.~\ref{fig:coordinates}: the vertical coordinate $z$ (m) is measured upward from the mean waterline ($z = 0$, positive upward), and the horizontal erosion depth $X(z)$ (m) accumulated over one wave of period $T$ is $X = V_m(z)\cdot T$.

\subsection{Spectral extension to irregular seas}
\label{sec:irregular}

Ocean waves are inherently irregular and stochastic. The sea state is described by an energy spectrum; we use a JONSWAP spectrum with peak-enhancement factor $\gamma=1$, which is mathematically equivalent to the Pierson--Moskowitz form applicable to fully developed seas:
\begin{equation}
S(\omega) = \frac{\alpha_{\mathrm{PM}} g^2}{\omega^5}
            \exp\!\left[-\frac{5}{4}\left(\frac{\omega_p}{\omega}\right)^4\right],
\label{eq:JONSWAP}
\end{equation}
where $\omega_p = 2\pi/T_p$ is the peak angular frequency and $\alpha_{\mathrm{PM}} = 8.1\times10^{-3}$ is the equilibrium-range constant. This spectrum is fully characterised by significant wave height $H_s$ and peak period $T_p$.

A common practice in wave-erosion modelling is to substitute $H_s$ and $T_p$ directly into Eq.~\eqref{eq:Vm_full}, treating the sea as a single regular wave. Since $H_s$ and $T_p$ are typically available only on an hourly basis, sub-hour wave irregularities are commonly left unaccounted for in wave-driven erosion studies \citep[e.g.,][]{barnhart2014}. Because the melt closure is nonlinear, evaluating it once at $(H_s,T_p)$ is not equivalent to averaging its response over the wave population. For the fixed $\gamma=1$ Pierson--Moskowitz spectral shape and power-law closure adopted here, nondimensionalisation makes the ensemble correction independent of $H_s$ and $T_p$: the component-wise treatment gives approximately 14\% less non-breaking erosion than the bulk-parameter treatment. Appendix~\ref{app:equivalence} derives this scale invariance and verifies the zero-upcrossing procedure independently against an analytical velocity-spectrum calculation.

A more rigorous treatment replaces the bulk-parameter approach with a \emph{wave-by-wave decomposition}. We define $X_{r,\mathrm{nb}}$ as the accumulated erosion profile assuming regular waves ($H_s, T_p$) without breaking, and $X_{ir,\mathrm{nb}}$ as the accumulated profile counting all individual waves in the time domain without breaking. 

For each hourly record $(H_s, T_p)$, a one-hour surface elevation time series $\eta(t)$ is generated from Eq.~\eqref{eq:JONSWAP} by assigning independent random phases and amplitudes $a_n=\sqrt{2S(\omega_n)\Delta\omega}$ for $n=1,\ldots,N_f$, where $\Delta\omega$ is the frequency-grid spacing and $N_f$ is the number of components. The spectrum is normalised so that $m_0\approx\sum_{n=1}^{N_f}a_n^2/2=H_s^2/16$; changing $N_f$ changes the discretisation, not the target variance. This irregular wave train is decomposed into individual waves by zero-upcrossing analysis, yielding a population $\{H_j, T_j\}_{j=1}^{N_w}$ for that hour. The wave count $N_w$ is set by the zero upcrossings of the generated record and is unrelated to the number of spectral components $N_f$. Using the deep-water form $R_0(z) = e^{k_j z}$ with $k_j = (2\pi/T_j)^2/g$, the per-wave (i.e., the $j$th wave) non-breaking erosion (where $m$ stands for melt) at depth $z$ is
\begin{equation}
X_{m,j}^{\mathrm{nb}}(z)
= C_{\mathrm{nb}}\,\frac{H_j}{T_j}\left(\frac{k_s}{H_j}\right)^{0.2}\!\Delta T_{\mathrm{wi}}\,T_j
  \cdot e^{k_j z},
\label{eq:Xm_nb}
\end{equation}
where $C_{\mathrm{nb}} = 2.09\times10^{-4}$~K$^{-1}$ is the Paper~1 baseline coefficient. The cumulative irregular non-breaking erosion profile is the summation over all waves across the entire observation period:
\begin{equation}
X_{ir,\mathrm{nb}}(z) = \sum_{\mathrm{all~hours}} \sum_{j=1}^{N_w} X_{m,j}^{\mathrm{nb}}(z).
\label{eq:X_ir_nb}
\end{equation}
Similarly, the regular non-breaking profile $X_{r,\mathrm{nb}}(z)$ is defined by summing the hourly contribution of a single representative wave over the same period:
\begin{equation}
X_{r,\mathrm{nb}}(z) = \sum_{\mathrm{all~hours}} C_{\mathrm{nb}}\,\frac{H_s}{T_p}\left(\frac{k_s}{H_s}\right)^{0.2}\!\Delta T_{\mathrm{wi}}\,(3600\,\mathrm{s}) \cdot e^{k_0 z}.
\label{eq:X_r_nb}
\end{equation}
At the waterline ($z = 0$), the period $T_j$ cancels between the orbital-velocity factor $H_j/T_j$ and the per-wave exposure duration $T_j$, giving
\begin{equation}
X_{m,j}^{\mathrm{nb}}\big|_{z=0}
= C_{\mathrm{nb}}\,k_s^{0.2}\,H_j^{0.8}\,\Delta T_{\mathrm{wi}},
\label{eq:Xm_Tcancel}
\end{equation}
which is independent of period. Appendix~\ref{app:equivalence} shows how the analytical velocity-spectrum and numerical zero-upcrossing routes both proceed from heat flux to erosion, and documents the compact power-law form used for the field calculation.

\subsection{Breaking-aware profile}
\label{sec:breaking}

Repeated front-collapse events leave behind a submerged ice foot that creates a shallowing region beneath the ice-shelf front (Fig.~\ref{fig:erosion_schematic}). As a wave with deep-water height $H_0$, period $T_0$, and wavelength $L_0 = gT_0^2/(2\pi)$ propagates over this ramp, it shoals and may break. Both transformations are governed by two standard functions of the dimensionless depth $d/L$ \citep{holthuijsen2007}: the shoaling coefficient $K_s = H/H_0$ and the wavelength ratio $K_L = L/L_0$, defined by
\begin{align}
K_s(d/L) &= \sqrt{\frac{1}{2n\,\tanh(kd)}}, \quad
n = \tfrac{1}{2}\!\left(1+\frac{2kd}{\sinh(2kd)}\right),
\label{eq:Ks_dL} \\
K_L(d/L) &= \frac{L}{L_0} = \tanh(kd).
\label{eq:KL_dL}
\end{align}
Here $K_s$ follows conservation of one-dimensional wave-energy flux for normal incidence. The formulation excludes bottom friction, dissipative losses before breaking, refraction, reflection, and three-dimensional spreading.

A direct implementation is expensive. For every wave and every depth step, the dispersion relation must be solved iteratively before the local wave height (Eqs.~\eqref{eq:Ks_dL}--\eqref{eq:KL_dL}) and the McCowan breaking check $H/d < 0.78$ can be evaluated. Applied to the roughly $7\times10^4$ individual waves in a two-month record, this brute-force approach would require hours of computation.

The key insight is that since $K_s$ and $K_L$ are universal functions of $d/L$ alone, the breaking condition $H_b = 0.78\,d_b$ combined with $H_b = H_0 K_s(d_b/L_b)$ reduces to a single equation linking two dimensionless quantities --- the deep-water steepness $H_0/L_0$ and the dimensionless breaker depth $d_b/L_b$:

\begin{equation}
\frac{H_0}{L_0} = 0.78\,\frac{d_b}{L_b}
\sqrt{\!\left(1 + \frac{4\pi\,d_b/L_b}{\sinh(4\pi\,d_b/L_b)}\right)
      \tanh^3\!\!\left(2\pi\frac{d_b}{L_b}\right)}.
\label{eq:lookup_master}
\end{equation}

Equation~\eqref{eq:lookup_master} is a universal relationship: for a given deep-water steepness $H_0/L_0$, there is a unique $d_b/L_b$. This relationship is pre-computed once and stored as a lookup table. For each wave, we then read $d_b/L_b$ from the table given its deep-water steepness and recover $L_b = K_L(d_b/L_b)\cdot L_0$, $d_b = (d_b/L_b)\,L_b$, $k_b = 2\pi/L_b$, and $H_b = 0.78\,d_b$ directly. A single array interpolation replaces the iterative dispersion calculation for every depth and wave, which makes the complete two-month record computationally practical.

\paragraph*{Erosion profile at breaking.} With $H_b$, $d_b$, $k_b$ known, the general forms (Eqs.~\eqref{eq:Rz_gen}--\eqref{eq:Rea_gen}) are evaluated at the breaking point:
\begin{equation}
R_b(z) = R(z;\,k_b,\,d_b) = \frac{\cosh[k_b(z+d_b)]}{\sinh(k_b d_b)},
\label{eq:Rz}
\end{equation}
\begin{equation}
\mathrm{Re}_b(z) = \frac{\pi H_b^2}{2 T_j \nu}\cdot R_b(z)^2.
\label{eq:Reb}
\end{equation}
Substituting into Eq.~\eqref{eq:Vm_full} gives the breaking-aware erosion depth per wave:
\begin{equation}
X_{m,j}^{b}(z)
= \frac{H_b}{T_j}
  \cdot
  \frac{0.0027\left(k_s/H_b\right)^{0.2} R_b(z)^{0.8}}
       {\mathrm{Re}_b(z)^{0.1}
        + \dfrac{3.84}{\sqrt{2}}\!\left(\mathrm{Pr}^{0.68}-1\right)}
  \,\Delta T_{\mathrm{wi}}\,T_j.
\label{eq:Xm_b}
\end{equation}
The cumulative irregular breaking-aware erosion profile is the summation over all waves:
\begin{equation}
X_{ir,b}(z) = \sum_{\mathrm{all~hours}} \sum_{j=1}^{N_w}
\begin{cases}
X_{m,j}^{b}(z) & z \in [0, -d_{b,j}] \\
X_{m,j}^{\mathrm{nb}}(z) & z < -d_{b,j}
\end{cases}.
\label{eq:X_ir_b}
\end{equation}

The combined profile applies an overwrite rule: the non-breaking profile (Eq.~\ref{eq:Xm_nb}) is retained for $z < -d_b$, while the shoaling zone $z\in[0,-d_b]$ is replaced by $X_{m,j}^b(z)$. Retaining the non-breaking profile below the breaker depth is a conservative lower bound, as breaking-induced turbulence likely penetrates somewhat deeper.

The McCowan breaking criterion ($H_b/d_b = 0.78$) is an idealised upper-bound parameterisation for the present geometry. Direct temperature- and salinity-induced changes to this inviscid ratio are secondary at the model order considered here, whereas local slope, reflection, ice-foot geometry, and dissipation before breaking may materially alter the breaking location.

In total, four model tiers are evaluated, each progressively relaxing simplifying assumptions:
\begin{itemize}
\item \textbf{T1 --- Regular, deep-water, non-breaking.} Each hour is represented by a single regular wave with bulk parameters $H_s$ and $T_p$. Evaluates $X_{r,\mathrm{nb}}$ (Eq.~\ref{eq:X_r_nb}). The simplified deep-water form ($e^{k_0 z}$) is used; no shoaling or breaking is considered. Sub-hour wave irregularities are not accounted for.
\item \textbf{T2 --- Irregular, deep-water, non-breaking.} Sub-hour wave irregularities are introduced through zero-upcrossing decomposition of a generated JONSWAP wave record. Evaluates $X_{ir,\mathrm{nb}}$ (Eq.~\ref{eq:X_ir_nb}). However, the simplified deep-water form is retained; no shoaling or breaking. This is a one-step spectral correction to T1.
\item \textbf{T3 --- Regular, breaking-aware.} Each hour is represented by a single regular wave with $H_s$ and $T_p$. Evaluates $X_{r,b}$. No sub-hour irregularities, but the depth-aware full-form is used: shoaling and breaking are considered, the water depth and erosion-profile depth are accounted for, and the breaking-aware erosion profile replaces the non-breaking profile within the shoaling zone. Each hour then contributes $3600\,\mathrm{s}/T_p$ identical waves, so the cumulative profile is $X_{r,b}(z) = \sum_{\text{all hours}} (3600\,\mathrm{s}/T_p)\, X_{m}^{b}(z)$ evaluated at $H_0 = H_s$ and $T_0 = T_p$ --- the breaking-aware analogue of Eq.~\eqref{eq:X_r_nb}, without the zero-upcrossing decomposition.
\item \textbf{T4 --- Irregular, breaking-aware.} Combines the spectral decomposition of T2 with the breaking-aware correction of T3. Evaluates $X_{ir,b}$ (Eq.~\ref{eq:X_ir_b}). This is the most complete tier considered under the stated linear-wave-theory assumptions.
\end{itemize}

\subsection{Field site and forcing data}
\label{sec:site}

\begin{figure}[t]
\centering
\includegraphics[width=0.85\textwidth]{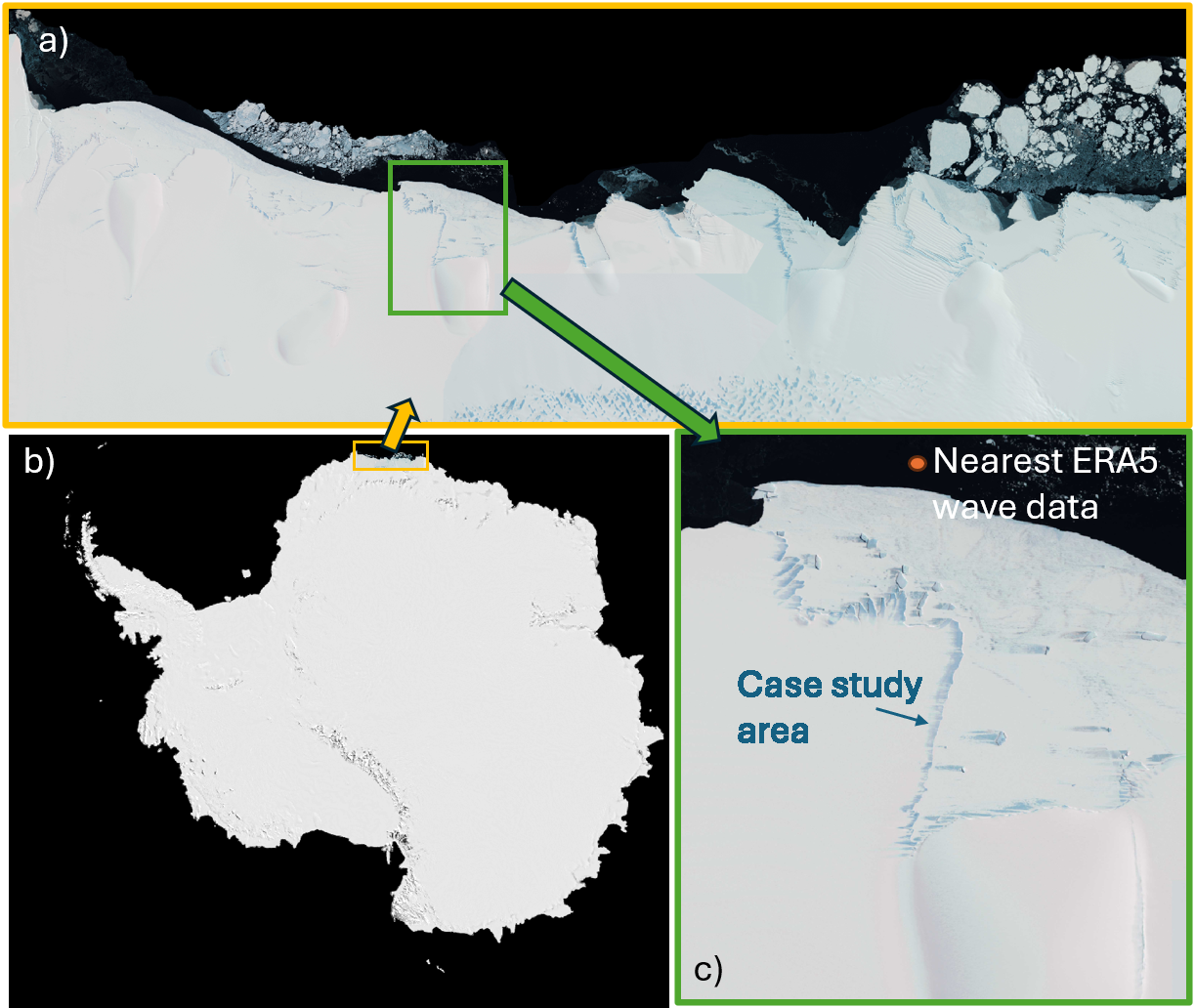}
\caption{Location of the study site. (a)~Satellite overview of
the Fimbulisen calving front; the green rectangle delineates the region shown in panel~(c).
(b)~Antarctica overview; the yellow marker indicates
Fimbulisen on the Dronning Maud Land coast, East Antarctica.
(c)~Close-up of the case study area; the orange circle marks the nearest ERA5
wave-reanalysis grid point (latitude $70.0^\circ$\,S, longitude $4.0^\circ$\,E,
approximately 23~km seaward of the calving front). Basemap data from
Quantarctica \citep{matsuoka2021quantarctica}.}
\label{fig:roi_location}
\end{figure}

The theoretical model tiers are applied to Fimbul Ice Shelf (Fimbulisen), a medium-sized ice shelf ($\sim$80\,km wide) located along Dronning Maud Land, East Antarctica, facing the open Weddell Sea. Its northern front is wave-exposed for much of the austral summer. We focus on the period January--February 2024, for which both continuous ERA5 wave reanalysis \citep{hersbach2020era5} and Sentinel satellite imagery are available. The exact study area and the location of the ERA5 forcing point are shown in Fig.~\ref{fig:roi_location}.

Offshore wave conditions are taken from the ERA5 hourly reanalysis at the grid point nearest to the ice front (latitude $70.0^\circ\,\mathrm{S}$, longitude $4.0^\circ\,\mathrm{E}$), providing significant wave height $H_s$ and peak period $T_p$. A dedicated wave-propagation study using detailed numerical modelling has shown that wave energy from this ERA5 grid point propagates to the calving front largely unchanged, owing to the deep-water conditions along the propagation path \citep{RN1854}. The waters off Fimbulisen are also largely free of sea ice in this season, so the open-ocean wave field reaches the front with little attenuation --- a further reason for selecting the January--February 2024 window \citep{RN1854}. The wave forcing proxy used throughout this study is the ratio $H_s/T_p$ (units m\,s$^{-1}$), which is directly proportional to the characteristic horizontal orbital velocity at the surface in the deep-water limit applicable to the ERA5 open-ocean grid point; the finite-depth correction $\coth(kd)$ is applied explicitly in Eq.~\eqref{eq:um} during the model calculation.

The thermal driving is set to $\Delta T_{\mathrm{wi}} = 1$~K, representing the difference between near-face water temperature and the local ice--water interfacial equilibrium temperature. At typical open-ocean salinity and near-atmospheric pressure, the seawater freezing temperature is approximately $-1.9\,^\circ$C. Under net melting, fresh meltwater lowers the interfacial salinity and raises the interfacial equilibrium temperature, while latent-heat consumption and cold meltwater can cool the adjacent water \citep{holland1999,mcconnochie2017}. Both effects reduce $\Delta T_{\mathrm{wi}}$ relative to a simple difference between regional sea-surface temperature and $-1.9\,^\circ$C. The 1~K value is therefore adopted as a physically plausible near-face baseline consistent with regional shelf-water conditions \citep{zhou2014asw}. Predicted waterline erosion scales linearly with $\Delta T_{\mathrm{wi}}$; direct near-front temperature and salinity profiles would refine its magnitude. The ice-surface roughness length is $k_s = 0.01\,\mathrm{m}$, adopted from \citet{white1980} and retained for consistency with Paper~1. All other physical constants are listed in Table~\ref{tab:constants}.

\subsection{Satellite observation of coastal retreat}
\label{sec:satellite}

Two complementary satellite products are used to derive the observed calving-front position time series. Sentinel-1 SAR (hereafter S1) provides frequent, all-weather observations from which ice-front polylines are manually digitised. Sentinel-2 optical imagery (hereafter S2) offers higher spatial resolution (10\,m vs.\ $\approx$\,40\,m for S1 EW mode) but is cloud-limited; front positions are extracted automatically using an S1-guided near-infrared gradient algorithm. The distinct radar and optical sensor responses provide a useful cross-sensor comparison, but the S2 retrieval is not statistically independent of the S1 reference geometry.

\subsubsection{Coastal line detections}
\label{sec:detections}

\textbf{Sentinel-1 dataset.}
Seventeen Sentinel-1 Extra Wide swath (EW) SAR acquisitions (ground resolution $\approx$\,40\,m) are available over Fimbulisen between 2~January and 28~February~2024, with a nominal repeat interval of approximately 3\,days. Ice-front polylines were manually digitised from each scene and stored as projected shapefiles in a polar stereographic coordinate system (EPSG:3031). The manual digitisation follows standard NPI protocols, tracking the radar-backscatter boundary between the ice shelf and open water (see Fig.~\ref{fig:coastlines_all} for the complete time series). Such delineations remain sensitive to ambiguous margins produced by sea ice and related backscatter features \citep{baumhoer2019}.

\textbf{Sentinel-2 dataset (Plateau-break detection).}
Five cloud-free Sentinel-2 scenes (7, 8, 17, and 27~January and 27~February; 10\,m resolution, EPSG:3031) are available over the study period. Rather than detecting each front from scratch, the search is anchored to the manually digitised S1 coastline from a neighbouring day.

This guidance absorbs two sources of ambiguity: the S1 delineations themselves carry uncertainty where storms or brash ice obscure the true ice face, and the front can shift between the S1 and S2 acquisition days --- retreating through sudden collapse or advancing with the background ice flow.

\begin{figure}[t]
\centering
\includegraphics[height=0.64\textheight,keepaspectratio]{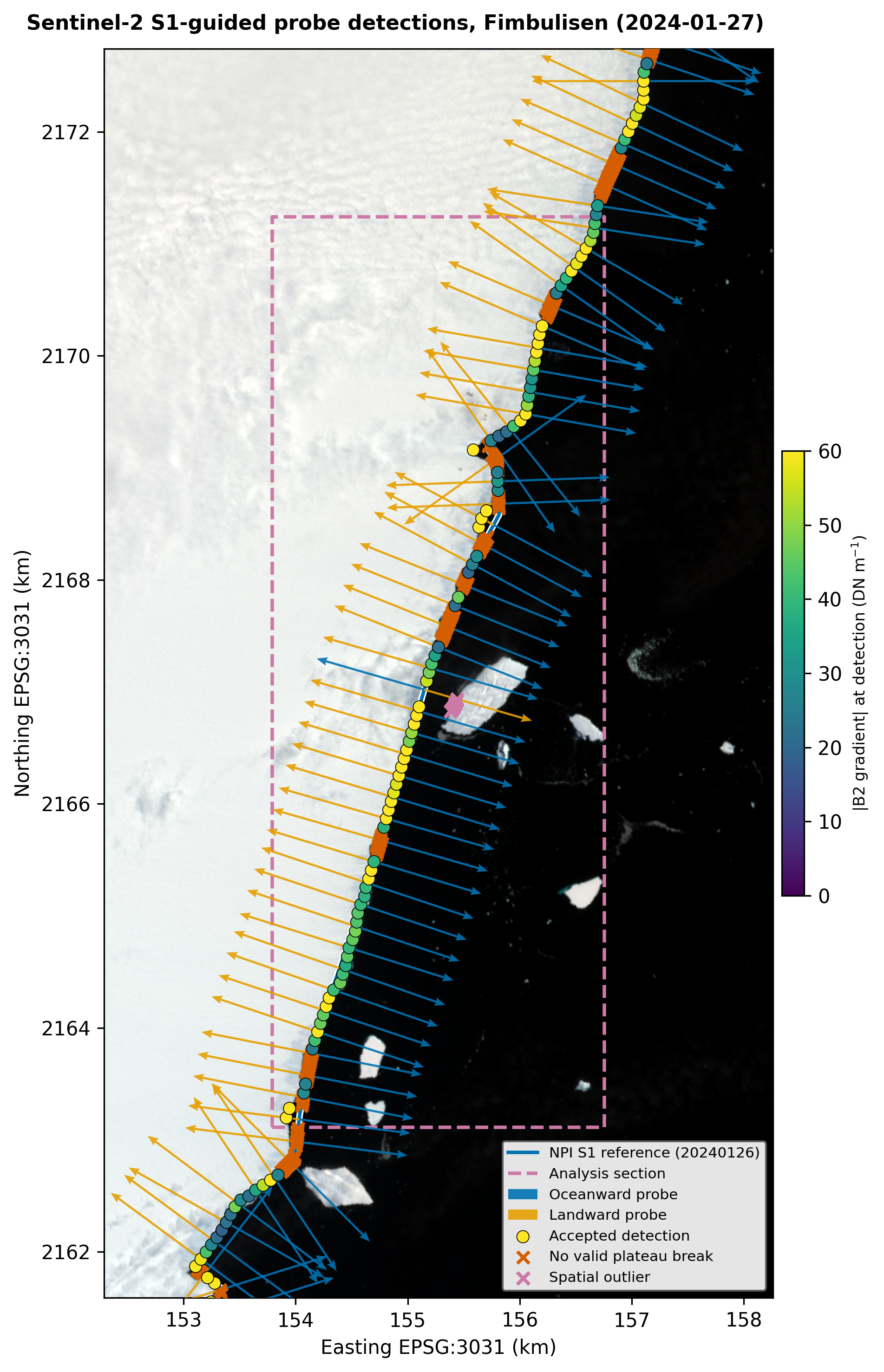}
\caption{S1-guided plateau-break probe geometry for the 27~January 2024 Sentinel-2 scene. Bidirectional probes are cast normal to the reference Sentinel-1 front to sample the NIR (B8) profile. Accepted detections (circles) are coloured by the magnitude of the NIR gradient; orange crosses mark seeds for which no valid plateau break was found, and purple crosses mark detections rejected as spatial outliers. Consecutive rejected markers indicate a contiguous reach with ambiguous NIR contrast or front geometry and are not interpreted as the length of a physical collapse event.}
\label{fig:s2_probe}
\end{figure}

From seed points along the S1 reference line, the algorithm probes the near-infrared (B8) image in both the seaward and landward directions and places the front where the bright ice-shelf plateau breaks abruptly to dark open-water values (Fig.~\ref{fig:s2_probe}). Detections that disagree strongly with their neighbours are rejected, and each accepted point carries a gradient-based confidence measure. Comparison against the S1 reference shows near-zero median offsets for the January scenes. The probe geometry, detection thresholds, fallback rules, and the station-level comparison with S1 are given in Appendix~\ref{app:s2_detection}; the implementation is included in the released code.

\subsubsection{Translating coastal line time series to coastal retreat}
\label{sec:retreat_methods}

Extracting a meaningful retreat signal from successive irregular coastal-line detections is non-trivial: the front geometry varies along-section, detections from different dates are not point-registered, and we wish to report both a section-mean and its spatial variability. We apply two complementary methods to each of the two sensor datasets.

\textbf{Method~1: Adaptive perpendicular (chainage).}
Separate reference coastlines are used for the two products: 2~January~2024 for S1 and 7~January~2024 for S2. Each reference is resampled at 50~m spacing and outward unit normals are computed at approximately 180 stations along the curved front, whose sampled chainage is approximately 8.9~km. For each subsequent detected front, the signed retreat at each station is the projection of the station-to-nearest-point vector onto the stored normal. This baseline-and-transect construction follows the general boundary-change framework used by the Digital Shoreline Analysis System \citep{himmelstoss2018}; the adaptive-normal and nearest-point implementation used here is study-specific. Cumulative retreat at each station is the running sum over time; the spatial mean and standard deviation across all stations define the reported band. This method preserves local spatial information but requires a stable reference geometry.

\begin{figure}[t]
\centering
\includegraphics[width=\textwidth]{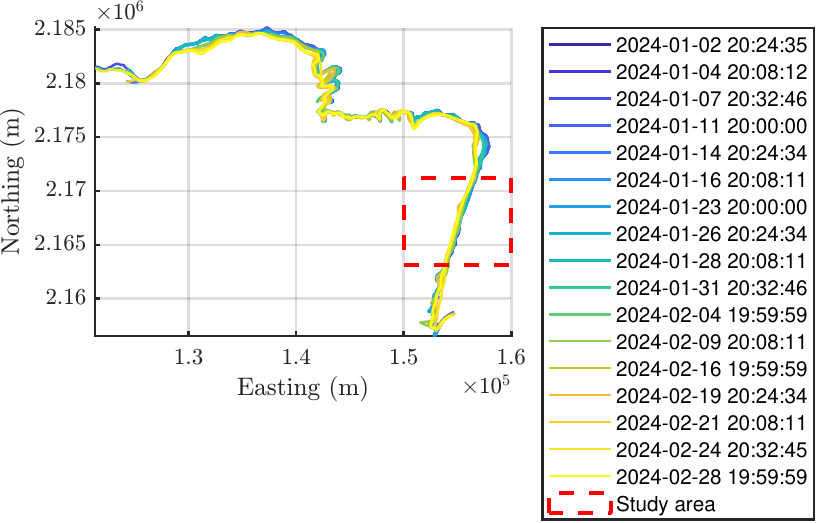}
\caption{Manually digitised Sentinel-1 coastal lines at Fimbulisen, coloured from earliest
(blue, 2 January 2024) to latest (yellow, 28 February 2024). The red dashed rectangle
delineates the analysis window, which spans 8.13~km in Northing
($N\in[2\,163\,110,\,2\,171\,240]$~m, EPSG:3031). The corresponding sampled
front chainage is approximately 8.9~km.}
\label{fig:coastlines_all}
\end{figure}

\textbf{Method~2: Area-based.}
For each consecutive front pair, both coastlines are clipped to the selected Northing range ($N\in[2\,163\,110,\,2\,171\,240]$~m; see Fig.~\ref{fig:coastlines_all}), and a closed polygon is formed by traversing the first coastline from south to north and the second from north to south. For its ordered vertices $(x_i,y_i)$, the enclosed area is calculated using the standard polygon-area formula:
\begin{equation}
A=\frac{1}{2}\sum_{i=1}^{M}\left(x_i y_{i+1}-x_{i+1}y_i\right),
\qquad (x_{M+1},y_{M+1})=(x_1,y_1).
\label{eq:shoelace}
\end{equation}
The nominal retreat is
\begin{equation}
\bar{X} = \frac{A}{L},
\label{eq:area_based}
\end{equation}
where $L$ is the straight-line distance between the endpoints of the first front in each pair (approximately 8.5~km for these data). Polygon orientation is selected so that $A/L>0$ denotes retreat and $A/L<0$ denotes advance. For a straight section undergoing uniform cross-shore retreat, $\bar{X}=A/L$ is equivalent to the chainage mean. This method suppresses local spatial variability but is less sensitive to individual front irregularities.

The analysis window was selected because the front is geometrically quasi-straight (Fig.~\ref{fig:coastlines_all}), which avoids the complications of along-section wave-propagation variability and ensures that the area-based and chainage estimators are well-defined. Its Northing span (8.13~km), sampled front chainage (8.9~km), and pair-specific area-normalisation length (approximately 8.5~km) describe different geometric quantities. Wave-induced melting is continuous whereas front collapse is episodic, so direct event-by-event comparison between modelled notch erosion and satellite-observed retreat is not straightforward. A longer observation window reduces sensitivity to the timing of individual collapse events, but it does not imply one-to-one convergence between the two quantities: cumulative retreat per unit notch erosion may exceed unity and may vary with fracture state, overhang geometry, and the evolving submerged ice foot. We therefore use the cumulative comparison to quantify the predictive shortfall of the erosion model, not as a direct validation of notch depth.

Both methods are applied to the S1 and S2 detected front time series, yielding four complementary retreat estimates in total.

\section{Results}
\label{sec:results}

\subsection{Wave forcing record}
\label{sec:waves}

\begin{figure}[t]
\includegraphics[width=\textwidth]{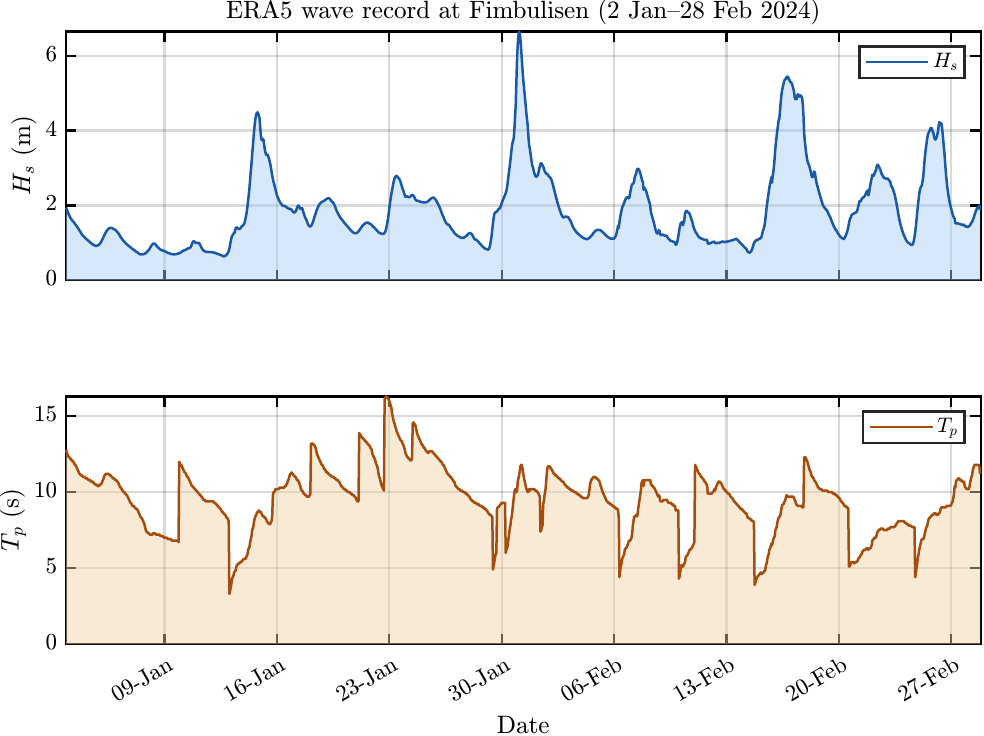}
\caption{ERA5 wave forcing at Fimbulisen over the Sentinel-1 observation window,
2~January--28~February 2024. Top: significant wave height $H_s$.
Bottom: peak period $T_p$.}
\label{fig:wave_history}
\end{figure}

Figure~\ref{fig:wave_history} shows the ERA5 $H_s$ and $T_p$ record at the Fimbulisen grid point over the Sentinel-1 observation window, 2~January--28~February 2024. Several energetic wave events with $H_s > 3$~m and $T_p > 12$~s occur during this interval. The wave forcing proxy $H_s/T_p$ shows strong episodic peaks.

\subsection{Event-scale comparison of wave forcing and retreat}
\label{sec:correlation}

\begin{figure}[t]
\includegraphics[width=\textwidth]{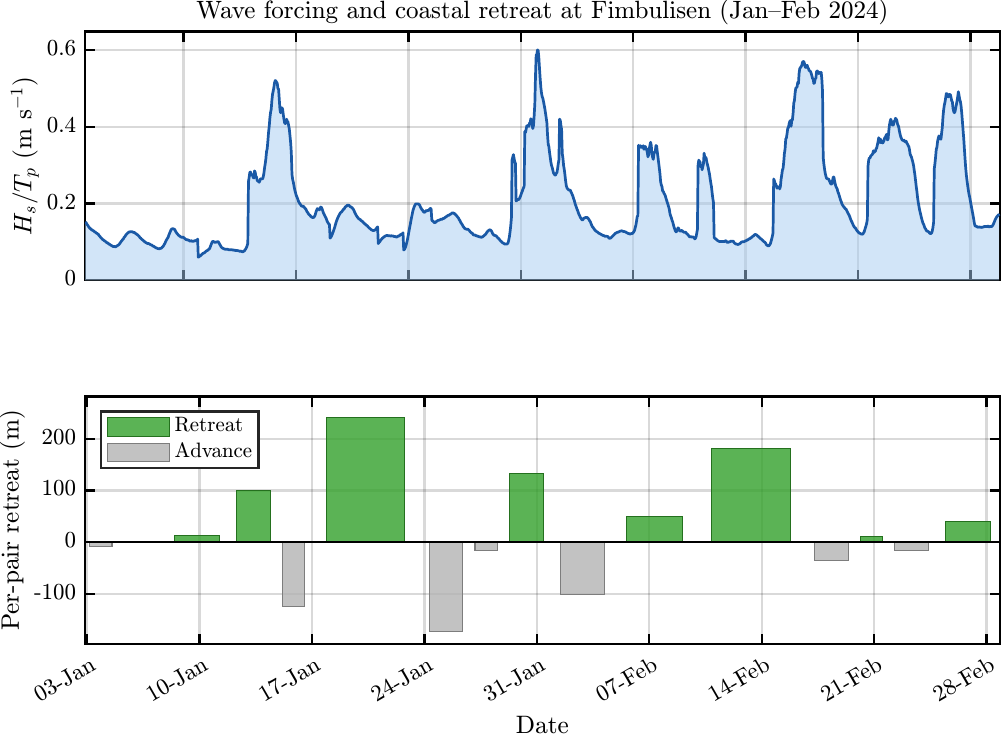}
\caption{Wave forcing and coastal retreat at Fimbulisen, January--February 2024.
Top: ERA5 wave forcing proxy $H_s/T_p$ (m\,s$^{-1}$).
Bottom: per-pair Sentinel-1 area-based retreat (green bars, retreat; grey bars, advance).
Bar width equals 70\% of the interval between consecutive Sentinel-1 acquisitions
(nominally $\approx 3$\,days); bars are centred at the acquisition midpoint.}
\label{fig:wave_retreat_correlation}
\end{figure}

Figure~\ref{fig:wave_retreat_correlation} overlays the ERA5 wave forcing proxy ($H_s/T_p$) with the per-pair Sentinel-1 area-based retreat. The incremental S1 record is highly scattered and includes occasional large negative (advancing) values. Continuous seaward ice flow contributes a real advancing component, but changes of this magnitude over approximately 3~days are unlikely to represent front motion alone. They probably combine true motion with delineation uncertainty associated with the $\approx$40\,m SAR imagery and ambiguous ice-margin features \citep{baumhoer2019}.

The per-pair S1 signal is therefore not a reliable indicator of individual storm-driven collapse events: for example, the largest observed increment (240~m between 16--23~January) occurs mainly during relatively calm conditions ($H_s/T_p \leq 0.2$~m\,s$^{-1}$), after a preceding storm peak ($H_s/T_p \approx 0.5$~m\,s$^{-1}$ around 17~January). Whether the preceding storm contributed to that later displacement cannot be determined from the per-pair record alone. We therefore compare model and observations in a cumulative, statistically averaged sense (Sect.~\ref{sec:model_comparison}), where short-term noise tends to cancel, consistent with Sect.~\ref{sec:retreat_methods}.

\subsection{Depth-resolved erosion profiles}
\label{sec:profiles}

\begin{figure}[t]
\includegraphics[height=0.64\textheight,keepaspectratio]{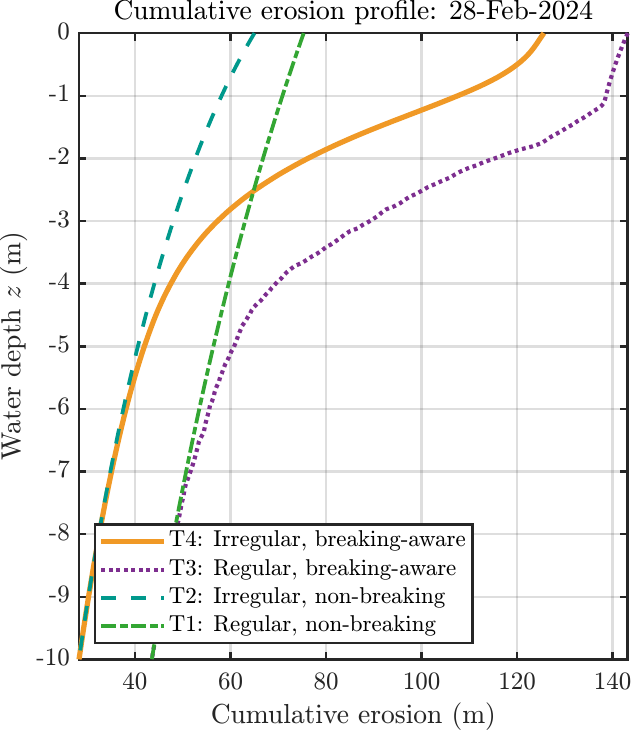}
\caption{Cumulative depth-resolved erosion profiles at the end of the
observation period (28 February 2024) for all four model tiers:
T4 irregular breaking-aware (orange, solid); T3 regular breaking-aware (purple, dotted);
T2 irregular non-breaking (teal, dashed); T1 regular non-breaking (green, dash-dot).
All profiles show maximum erosion at the waterline ($z=0$) decaying with depth.
The enhanced undercutting within the notch zone ($z\in[0,-d_b]$) is visible
in the breaking-aware tiers (T3, T4).}
\label{fig:profiles_final}
\end{figure}

Figure~\ref{fig:profiles_final} shows the cumulative depth-resolved erosion profile at the end of the observation period for T1, T2, T3, and T4. All tiers show maximum erosion at the waterline ($z=0$), decaying with depth following the orbital velocity envelope. For T4, the breaking-aware correction amplifies erosion within the shoaling zone $z \in [0, -d_b]$ above the ice foot relative to the non-breaking baseline, creating a distinctly more undercut notch profile. This morphology is consistent with field observations of thermo-erosional notch geometry at calving fronts \citep{roehl2006}, and provides the physical basis for the shoaling-amplified forcing that distinguishes T3 and T4 from their non-breaking counterparts.

\subsection{Model tiers and satellite observations}
\label{sec:model_comparison}

\begin{figure}[t]
\includegraphics[width=\textwidth]{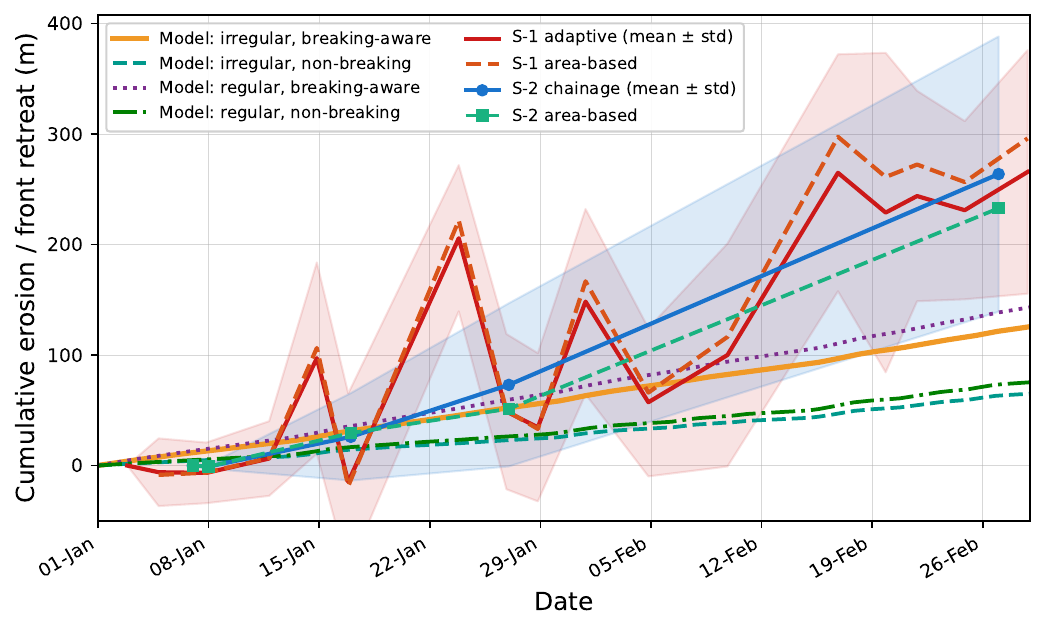}
\caption{Comparison of modelled cumulative waterline erosion and satellite-observed
front retreat at Fimbulisen, January--February 2024.
Four model tiers: T1 regular non-breaking (green, dash-dot);
T2 irregular non-breaking (teal, dashed);
T3 regular breaking-aware (purple, dotted);
T4 irregular breaking-aware (orange, solid).
Here ``regular'' and ``irregular'' denote the hourly bulk-wave and spectrally decomposed
model representations, respectively; they are not observed sea-state classifications.
Model curves show raw accumulation from 1~January; matched-window values quoted in the
text are obtained by subtracting the model value at the first satellite acquisition.
Observations: Sentinel-1 adaptive perpendicular mean $\pm$ 1 standard deviation
(red band and curve); Sentinel-1 area-based (orange dashed);
Sentinel-2 plateau-break chainage mean $\pm$ 1 standard deviation (blue band and curve);
Sentinel-2 area-based (teal dashed squares).}
\label{fig:main_comparison}
\end{figure}

Figure~\ref{fig:main_comparison} shows the cumulative waterline erosion predicted by the four model tiers alongside the satellite-observed front retreat over January--February 2024. The comparison is a one-to-one baseline (hereafter, the baseline comparison); the model does not simulate the fracture and calving processes that convert notch growth into front displacement.

Among the empirical LWT tiers, T1 (which applies bulk parameters $H_s$ and $T_p$ as a single regular wave each hour, i.e.\ ignoring sub-hour wave irregularities) produces more erosion than T2 (which evaluates the resolved wave population). T2 is approximately 14\% lower than T1, consistent with the scale-invariant correction derived for the adopted $\gamma=1$ Pierson--Moskowitz spectrum and power-law closure (Appendix~\ref{app:equivalence}). The breaking-aware correction (T3, T4) substantially elevates the prediction. For direct comparison with the Sentinel-2 chainage record, model erosion is differenced over the same 7~January--27~February interval. Over this matched window, the raw T4 accumulation increases from 12 to 122~m, giving approximately 110~m of incremental waterline erosion. T4 is the most complete tier considered under the stated LWT assumptions.

Over 7~January--27~February, the S2 plateau-break chainage method yields $264 \pm 125$~m (mean $\pm$ 1 s.d.\ across all stations); the S1 adaptive perpendicular mean reaches $266$~m over its slightly longer 2~January--28~February window. The section means are therefore similar, differing by about 2~m despite their different endpoints. This agreement and the large along-front spread ($\pm 125$~m) are assessed in Sect.~\ref{sec:obs_uncertainty}; neither is a direct accuracy estimate. The station-level retreat varies considerably, and at most stations the model predictions fall outside the observed range. Even the lowest area-based estimate ($233$~m for S2) far exceeds the modelled erosion. Comparing the matched T4 increment (110~m) with the S2 chainage estimate (264~m) gives a baseline shortfall factor of 2.4. The sensitivity of this result to the calibration coefficient $\alpha$ and the thermal driving $\Delta T_{\mathrm{wi}}$ is discussed in Sect.~\ref{sec:turbulence_gap}.

\section{Discussion}
\label{sec:discussion}

\subsection{Systematic bias from bulk-parameter substitution}
\label{sec:Hs_discussion}

Our results demonstrate that applying the White (1980) formula directly with $H_s$ and $T_p$ (tier T1) overestimates the melt prediction relative to the component-wise formulation (T2). The reason is structural: the melt rate scales approximately as $H/T$ with a sub-linear roughness correction $(k_s/H)^{0.2}$, so evaluating the closure at bulk parameters does not reproduce its wave-population average. The power-law friction approximation used at field scale remains close to the full Lambert-$W$ closure in the applicable excursion-to-roughness range; it therefore does not alter this interpretation. For the fixed $\gamma=1$ Pierson--Moskowitz spectral family adopted here, the nondimensional wave population is scale invariant and T2 is 14\% lower than T1, independent of $H_s$ and $T_p$ in the long-record limit. This percentage is specific to the adopted spectral shape and melt closure; spectra with different shape parameters or a different closure need not have the same correction.

\subsection{Breaking waves as a first-order correction}
\label{sec:breaking_discussion}

The breaking-aware correction (T3, T4) substantially increases the predicted waterline erosion through the enhanced orbital velocities at the shoaling limit above the ice-foot ramp. Physically, this represents the scenario where incoming waves propagate over the increasingly shallow ice foot, steepen, and approach the breaking condition before impacting the ice face. The velocity amplification factor, defined as the ratio of the breaking-aware orbital velocity at the waterline to the deep-water orbital velocity $\pi H_0/T$ of the same wave, can reach 1.5--2.0 for typical Fimbulisen conditions, roughly doubling the thermal forcing at the notch tip compared to the non-breaking estimate.

The breaking-aware correction is conditional on idealised propagation assumptions. It conserves one-dimensional wave-energy flux for normal incidence and uses $H_b/d_b=0.78$ as an upper-limit breaker index; bottom friction, pre-breaking dissipation, refraction, reflection, three-dimensional spreading, and detailed ice-foot slope effects are omitted. It therefore estimates the orbital velocities attainable under idealised shoaling but does not resolve the turbulent dissipation, air entrainment, undertow, or boundary-layer disruption following actual breaking. T3 and T4 should consequently be interpreted as breaking-aware extensions within LWT, rather than site-specific surf-zone models.

\subsection{Magnitude and sensitivity of the predictive shortfall}
\label{sec:turbulence_gap}

The quantitative comparison established in Sect.~\ref{sec:model_comparison} clarifies the structure of the wave-erosion problem. The White-lineage empirical framework predicts substantially more erosion than the molecular-diffusion streaming theory discussed below in Sect.~\ref{sec:laminar_discussion} because its Stanton-number closure implicitly absorbs turbulent boundary-layer enhancement from the laboratory calibration environment --- specifically, the turbulent (non-breaking) wave boundary-layer enhancement quantified by \citet{RN1842} --- even without explicitly modelling post-breaking turbulence. With spectral decomposition and shoaling-amplified breaking waves (T4), the baseline calculation still falls short of satellite observations by a factor of 2.1--2.4.

We reserve the term \emph{Turbulence Gap} for the unresolved hydrodynamic component of this shortfall; the distinct roles of hydrodynamic enhancement, fracture-controlled amplification, and thermal forcing are examined below.

\subsubsection*{Robustness of the shortfall estimate}

The predictive shortfall factor of $\approx 2.4$ is conditional on the adopted calibration coefficient and thermal driving. Their effects can be separated because the predicted erosion is proportional to both $\alpha$ and $\Delta T_{\mathrm{wi}}$ in the compact field-scale closure.

\textit{Calibration coefficient.} We adopt $\alpha = 1$ as the theoretical baseline. Applying the representative laboratory-calibrated value $\alpha = 0.648$ derived in Sect.~\ref{sec:governing} would reduce $C_{\mathrm{nb}}$ from $2.09\times10^{-4}$ to $1.35\times10^{-4}$~K$^{-1}$, lowering the T4 erosion prediction from $\approx110$~m to $\approx70$~m and widening the baseline shortfall to a factor of $\approx 3.7$.

\begin{figure}[t]
\centering
\includegraphics[width=\textwidth]{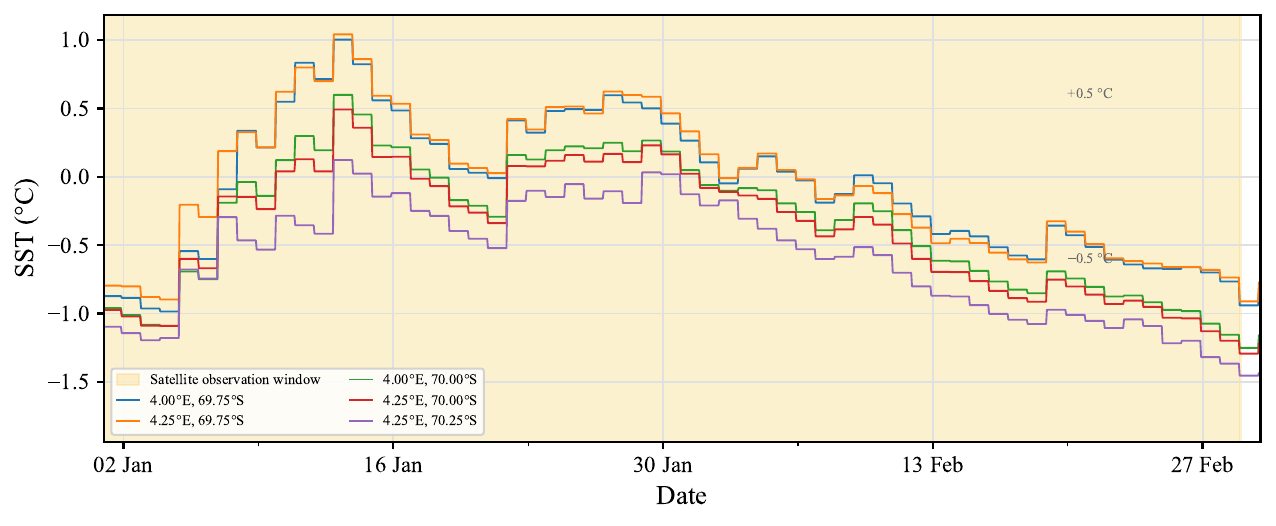}
\caption{ERA5 hourly sea-surface temperature at the five ERA5 grid points nearest the Fimbulisen
calving front (longitude $4.0^\circ$--$4.25^\circ$\,E, latitude $69.75^\circ$--$70.25^\circ$\,S),
January--March 2024. Near-front temperatures during the satellite observation window
(amber shading, January--February) are generally between $-0.5$ and $+0.5$\,$^\circ$C.
Relative to a representative near-surface seawater freezing temperature at ambient salinity of
$-1.9$\,$^\circ$C, these regional values imply thermal driving of approximately
$1.4$--$2.4$~K. The model adopts a mildly conservative near-face baseline of
$\Delta T_{\mathrm{wi}} = 1$~K; predicted waterline erosion varies linearly with this quantity.}
\label{fig:sst_temperature}
\end{figure}

\textit{Thermal driving.} We adopt $\Delta T_{\mathrm{wi}} = 1$~K as a physically plausible baseline \citep{zhou2014asw}. Nearby ERA5 sea-surface temperatures generally range from approximately $-0.5$ to $+0.5$\,$^\circ$C during the observation window (Fig.~\ref{fig:sst_temperature}). Relative to a representative near-surface seawater freezing temperature of $-1.9$\,$^\circ$C at ambient salinity, this range corresponds to regional surface thermal driving of approximately $1.4$--$2.4$~K. The upper values occur only briefly, while much of the record lies closer to the lower end. At a near-surface melting interface, fresh meltwater lowers the salinity relative to ambient seawater and therefore raises the interfacial equilibrium temperature above $-1.9$\,$^\circ$C \citep{holland1999,mcconnochie2017}. Heat consumed by melting and cold meltwater entering the near-face layer can also make the adjacent water cooler than the regional surface water. Both effects reduce the local thermal driving relative to the simple ERA5-minus-$(-1.9\,^\circ\mathrm{C})$ estimate. A baseline of 1~K is therefore a reasonable and mildly conservative representation of near-face conditions.

These thermodynamic effects support, rather than undermine, the 1~K baseline; it is not chosen to maximise the predicted erosion. The exact time-mean $\Delta T_{\mathrm{wi}}$ remains unmeasured and changes the predicted magnitude linearly, but it is a forcing uncertainty rather than a missing amplification mechanism. The calibrated value of $\alpha$ widens the shortfall at fixed $\Delta T_{\mathrm{wi}}$, whereas a larger time-mean thermal driving would reduce it proportionally. The factor of 2.4 is therefore a physically motivated baseline model--observation ratio rather than a universal constant. Within this interpretation, the leading unresolved process mechanisms remain hydrodynamic enhancement of notch erosion and fracture-controlled amplification from notch growth to front retreat. Direct near-ice profiles would tighten the quantitative partition without replacing either mechanism.

\subsection{Attributing the shortfall: hydrodynamic enhancement and fracture amplification}
\label{sec:hydro_vs_glacio}

Two unresolved processes may contribute to the predictive shortfall: post-breaking hydrodynamic enhancement of heat transfer at the ice face and fracture-controlled amplification between notch erosion and observable front retreat. The hydrodynamic processes determine how rapidly the notch develops, whereas fracture and collapse determine how much retreat is produced by a given notch depth. Uncertainty in thermal driving affects the magnitude of the modelled notch erosion but represents a forcing uncertainty rather than a separate amplification mechanism. The hydrodynamic candidates include wave-breaking turbulence and mixing at the ice face, undertow and return flow, notch-confined jetting, and intermittent run-up or slamming. Laboratory surf-zone measurements demonstrate that breaking can generate both strong turbulence and a structured offshore undertow \citep{ting1994}; analogous processes may enhance near-ice heat transfer beyond the non-breaking boundary-layer contribution represented by the Stanton-number closure \citep{RN1842}. These motions were not measured at Fimbulisen, so they are hypotheses for targeted laboratory and field tests rather than mechanisms established by the present observations.

Future experiments should test whether the near-front hydrodynamic regime evolves systematically with notch geometry: from predominantly reflective interaction at a near-vertical face, through ramp-like jetting and slamming as the notch develops, to wave propagation and breaking within a mature notch. This sequence is a working hypothesis rather than a process established by the present field observations.

Wave-energy dissipation is ultimately converted to heat, but any local temperature increase depends on the dissipation volume, residence time, mixing, and advection, and cannot be quantified from the available observations. At the model order considered here, the principal expected effect of breaking is enhanced mixing and interfacial heat transfer; direct local warming is not represented.

Real ice-shelf fronts are not pristine slabs. Pre-existing crevasse fields, basal fractures, and the continuous bending moment from the buoyant submarine ice foot \citep{wagner2014footloose} reduce the critical notch depth at which a slab detaches. In principle, this mechanical weakening could cause the front to fail at shallower notch depths and produce more retreat per unit of wave energy than a pristine-slab model predicts.

These two categories act on different physical quantities, but they do not simply compete --- they multiply. The wave-erosion model predicts the \emph{rate} at which the notch deepens: a continuous, wave-by-wave heat-transfer process accumulating over weeks. Fracture mechanics governs how much observable front retreat each increment of notch growth ultimately produces. For a pristine slab that calves back exactly to the notch tip, retreat equals erosion. Real fronts need not behave this way: front collapse removes the slab overlying the notch, and repeated collapse cycles develop the submerged ice foot whose uncompensated buoyancy can trigger footloose calving that detaches slabs whose dimensions are set by the flexural response of the ice shelf rather than by the notch depth itself \citep{wagner2014footloose,sartore2025}. At the conceptual level, the conversion may be written schematically as $X_{\mathrm{retreat}}=A X_{\mathrm{notch}}$, but $A$ is neither a constant nor identifiable from the present observations: it depends on the fracture threshold, timing within the erosion--collapse cycle, pre-existing weaknesses, and footloose amplification.

This distinction sharpens the interpretation of the shortfall: the satellite comparison constrains the \emph{product} of the unmodelled hydrodynamic enhancement and the mechanical amplification, not the Turbulence Gap alone. The comparison therefore cannot rank post-breaking hydrodynamic enhancement against mechanical amplification. Post-breaking turbulence remains physically plausible because an enhanced effective thermal diffusivity under breaking waves is expected beyond the non-breaking enhancement already encoded in the Stanton-number closure \citep{RN1842}. However, the sparse front-position record does not establish event-by-event attribution, and wave-triggered fracture could introduce a delayed or discrete response to the same forcing. Separating erosion from amplification requires observations that sense the hidden notch or its mechanical precursors. Candidates include near-front notch profiling, deformation records at the front, and acoustic signatures of individual collapse events. These observations must then be interpreted through coupled wave-erosion and fracture modelling. Reduced-order analysis of notch-induced overhang instability \citep{xu2026iahr} indicates that the fracture side of this coupling is analytically tractable. The coupling itself --- rather than either mechanism in isolation --- is the missing element of current ice-shelf erosion frameworks, and quantifying it is beyond the scope of the present study.

\subsection{Comparison with molecular-diffusion streaming theory}
\label{sec:laminar_discussion}

To place our empirical results in a broader theoretical context, we evaluate the first-principles boundary-layer streaming theory of \citet{wolterman2026} as an analytical baseline. Their parameter-free wave-induced melt formulation was developed for reflecting, non-breaking monochromatic waves and evaluated experimentally at a fixed angular frequency of $9.42$~rad~s$^{-1}$ ($T=0.667$~s). It does not represent wave breaking or the evolving ice-foot geometry considered here. Using the notation of this paper, its erosion rate at the waterline is
\begin{equation}
V_m^{\mathrm{W}}\big|_{z=0}
= \sqrt{\frac{6\alpha_{\mathrm{th}}\omega}{\pi}}\,\frac{\rho_w}{\rho_i}\,\frac{c_p\,\Delta T_{\mathrm{wi}}}{L_f}\,k\,a,
\label{eq:wolterman}
\end{equation}
where $\alpha_{\mathrm{th}}$ (m$^2$\,s$^{-1}$) is the thermal diffusivity of water, $\rho_w/\rho_i$ is the water-to-ice density ratio, $L_f$ (J\,kg$^{-1}$) is the latent heat of fusion, and $a = H/2$ (m) is the wave amplitude.

For the field comparison, Eq.~\eqref{eq:wolterman} is evaluated wave by wave using the same zero-upcrossing wave heights $H_j$ and periods $T_j$ extracted from the hourly ERA5-derived irregular-wave histories used in T2. For each wave, the contribution $V_{m,j}^{\mathrm{W}}T_j$ is accumulated over the matched 7~January--27~February interval. This gives approximately 2~m of cumulative waterline erosion, compared with 59~m from T2 and 110~m from T4.

The small Wolterman estimate follows from the molecular-diffusion streaming closure, for which the waterline melt rate scales as $a\omega^{5/2}\Delta T_{\mathrm{wi}}$. In the deep-water limit, the corresponding per-wave erosion scales as $X_j^{\mathrm{W}}\propto H_jT_j^{-3/2}\Delta T_{\mathrm{wi}}$, whereas the White-lineage T2 prediction scales as $X_{ir,\mathrm{nb}}\propto H_j^{0.8}$ (Eq.~\ref{eq:X_ir_nb}) and is independent of period at the waterline (Eq.~\ref{eq:Xm_Tcancel}). This period dependence is a physical prediction of the streaming theory and should not be interpreted simply as a laboratory-scale artefact. Indeed, Wolterman's experimental period of 0.667~s is comparable to the 0.87 and 1.54~s periods used in Paper~1. The more important limitation is that the theory was developed and evaluated for smooth, non-breaking conditions governed by molecular thermal diffusion, whereas the Fimbulisen application involves roughness, irregular seas, shoaling, breaking, and notch-confined flow.

A direct comparison with Paper~1 further illustrates this regime dependence. Evaluated with the measured incident-wave conditions and finite-depth wavenumbers, Eq.~\eqref{eq:wolterman} gives mean waterline melt rates of approximately 7.9 and 62.0~$\mu$m~s$^{-1}$ for Tests~1 and~2, respectively, compared with observed mean recession rates of approximately 54.7 and 108.2~$\mu$m~s$^{-1}$. This comparison is indicative rather than definitive because Paper~1 did not include a no-wave control, whereas \citet{wolterman2026} isolated the wave-induced contribution by subtracting ambient melting. The two studies also differ in relative depth, wave amplitude, reflection treatment, surface condition, and the intermittent ice-face impact observed in Paper~1 Test~2. Nevertheless, the comparison shows that the molecular streaming mechanism alone does not consistently reproduce the total erosion measured across the two laboratory conditions.

Wolterman's formulation therefore provides a valuable mechanistic baseline for the wave-induced component under idealised non-breaking conditions, but it does not close the Fimbulisen comparison. It does not represent rough-wall turbulent transfer, post-breaking turbulence, notch-confined flow, or the fracture-controlled amplification between hidden notch growth and satellite-observed retreat. The 2~m field estimate consequently supports the need to resolve additional hydrodynamic transport, but it cannot by itself partition the observed shortfall between enhanced notch erosion and fracture amplification.

\subsection{Satellite observation uncertainty and implications for model validation}
\label{sec:obs_uncertainty}

The two complementary satellite products yield mean section-level retreats of $266$~m (S1 adaptive perpendicular chainage) and $264$~m (S2 plateau-break chainage) over slightly different windows. Their similar section means are encouraging because radar backscatter and visible-band values respond to different physical contrasts at the calving face. They support the interpretation that both products track the same section-scale ice-front displacement, but the approximately 2~m difference is not a measurement-accuracy estimate: the S2 search is guided by S1 reference geometry, the endpoints differ, and neither product proves that every local detection excludes brash ice, sea ice, or other secondary margin features.

It is useful to distinguish the inter-sensor difference between section means from the along-front spread within either product. The chainage section means differ by $\approx 2$~m, whereas the S2 retreat values have an along-front standard deviation of $\pm 125$~m. This standard deviation is not a direct measurement-error estimate: it combines genuine spatial heterogeneity with local retrieval uncertainty. The S1 per-pair series also contains large apparent advances (Fig.~\ref{fig:wave_retreat_correlation}). Background ice flow can produce real advance, but the largest short-interval changes probably include substantial delineation uncertainty at the $\approx$40~m SAR resolution. We therefore emphasise section-mean cumulative displacement and retain the product spread when interpreting the model shortfall.

The area-based estimates diverge more ($296$~m for S1 vs $233$~m for S2, a difference of $\approx 63$~m), consistent with SAR and optical sensors delineating slightly different ice-margin features in complex sea-ice conditions. This product spread provides a practical measure of observational uncertainty for model validation. Even the lower S2 area-based estimate ($233$~m) exceeds the baseline T4 prediction by a factor of $\approx 2.1$, so the direction of the baseline shortfall is consistent across the two retrieval methods.

\subsection{Implications for ice-sheet models}
\label{sec:ism_implications}

Wave-induced erosion --- including thermo-erosional notch formation, waterline melting, and their potential contribution to front collapse --- is not represented explicitly in the major community ice-sheet model parameterisations considered here. Common continental-scale calving laws (eigencalving, von Mises stress, and crevasse-depth criteria) are governed by ice rheology and far-field stress rather than ocean-surface wave forcing \citep{RN1833}. Evaluation studies of these operational calving laws have not included wave-erosion schemes \citep{wilner2023}, and \citet{sartore2025} recently adapted an iceberg-decay wave-erosion formulation to estimate frontal ablation at the Ross Ice Shelf front. In the baseline calculation, the extended LWT erosion estimate is smaller than observed front retreat by a factor of 2.1--2.4, depending on the satellite product. Omitting wave-induced frontal erosion may therefore overestimate the stability of some wave-exposed ice shelves and underestimate their frontal mass loss, although the magnitude remains conditional on unresolved thermal driving and fracture amplification.

\subsection{Event-scale forcing and attribution}
\label{sec:event_attribution}

The per-pair retreat record (Fig.~\ref{fig:wave_retreat_correlation}) is episodic, but its short-interval variability cannot be matched uniquely to individual wave events because satellite acquisitions are sparse and the series includes apparent advances and delineation uncertainty. In the model, notch erosion accumulates continuously and nonlinearly from the full wave history, whereas observable retreat occurs through discrete fracture and collapse events. A cumulative wave-energy-flux diagnostic, such as $\int H_s^2 C_g\,\mathrm{d}t$, where $C_g$ is group velocity, could help test lagged associations over longer records, but it would not replace the wave-by-wave heat-transfer model or distinguish hydrodynamic enhancement from fracture amplification. Resolving event-scale causation requires denser, synchronised wave, thermal, notch-geometry, and front-position observations. More fundamentally, continuous notch growth and episodic front retreat cannot be compared one-to-one until the intervening erosion--fracture sequence is represented. A statistically stable cumulative relationship may emerge only after notch, overhang, and submerged-foot mechanics, together with a sufficient ensemble of erosion--collapse cycles, are included. Establishing whether such a conversion exists is therefore a testable hypothesis beyond the present data.

\section{Conclusions}
\label{sec:conclusions}

We evaluated the gap between theoretical wave hydrodynamics and observed ice-shelf front retreat by extending the empirical linear wave theory (LWT) framework and comparing it against satellite observations at the Fimbul Ice Shelf front during January--February 2024. Our primary contribution is an extension of the \citet{white1980} LWT formulation to irregular seas and breaking-aware wave profiles, evaluated against Sentinel-1 and Sentinel-2 satellite observations. The key conclusions are:

\begin{enumerate}

\item \textbf{Extended LWT accounts for part of the observed retreat in a one-to-one baseline comparison, but falls short.}
Incorporating spectral decomposition and a shoaling-amplified breaking-wave correction gives approximately 110~m of cumulative waterline erosion over the matched 7~January--27~February interval. The Sentinel-2 chainage observations record 264~m of front retreat over the same interval, giving a baseline shortfall factor of about 2.4. For the adopted $\gamma=1$ Pierson--Moskowitz spectrum and power-law closure, component-wise spectral treatment reduces the predicted erosion by 14\%, independent of $H_s$ and $T_p$, while the breaking-aware correction nearly doubles it. These two effects partially offset each other, and neither closes the model--observation gap.

\item \textbf{The baseline predictive shortfall reflects coupled missing physics and forcing uncertainty.} The extended LWT model (T4, irregular breaking-aware) falls short of the satellite estimates by a factor of 2.1--2.4, depending on the product; the 2.4 value follows from the common 7~January--27~February model--Sentinel-2 interval with $\alpha=1$ and $\Delta T_{\mathrm{wi}}=1$~$^\circ$C. Applying the laboratory-calibrated $\alpha$ widens this ratio, while the unmeasured near-ice thermal driving can shift it in either direction. The residual may combine post-breaking surf-zone hydrodynamics with fracture-controlled amplification, in which front collapse and footloose calving convert hidden notch erosion into larger observable retreat. Front-position observations alone cannot separate these factors. Given the large along-front variability and the spread between the (non-independent) satellite products, the comparison rests on section-mean cumulative retreat rather than on individual stations or image pairs.

\item \textbf{Event-scale attribution remains unresolved.} Modelled notch erosion accumulates continuously from the full wave history, whereas satellite-observed retreat occurs in discrete, sparsely sampled increments. The available record does not establish that the largest retreat increments coincide directly with the largest wave events; wave-induced mixing and fracture-controlled collapse may introduce delayed or amplified responses.

\end{enumerate}

The wave-erosion model code (Tiers T1--T4) uses a vectorised lookup-table scheme that avoids repeated iterative dispersion calculations and makes application to multi-month wave records practical. The model, satellite-analysis code, and processed results have been prepared for public archival with the manuscript.

While this study is specific to Fimbul Ice Shelf, the theoretical extensions are general: spectral decomposition and the breaking-aware shoaling profile can be applied to wave-exposed ice fronts with open-water fetch, subject to the stated propagation assumptions. Wave-induced frontal erosion is not represented explicitly in the major ice-sheet model parameterisations considered here. A conservatively parameterised erosion term could therefore provide a useful route towards representing this process at wave-exposed ice margins.

These findings motivate three lines of further work. Near-ice turbulence and thermal boundary-layer dynamics at wave-exposed fronts need dedicated laboratory and field quantification. Coupled wave-erosion and fracture modelling is needed to quantify the mechanical amplification between hidden notch erosion and observable retreat. Finally, higher-resolution SAR acquisitions and longer averaging windows would reduce the spatial variability of satellite-derived front retreat.

\appendix
\section{Equivalent-wave and zero-upcrossing treatments of irregular-wave heat flux}
\label{app:equivalence}

This appendix verifies the irregular-wave calculation by comparing two independent routes: an analytical equivalent-wave treatment based on the velocity spectrum and a numerical treatment based on a generated wave history and zero-upcrossing decomposition. Both routes calculate the heat flux first and then use the same Stefan condition to obtain melt rate and erosion distance. The comparison therefore tests the spectral treatment itself; the conversion from heat flux to erosion is common to both routes.

\subsection*{A1.\enspace Common heat-flux-to-erosion relation}

Let $q_w(t)$ be the heat flux from the water to the ice (W\,m$^{-2}$). Assuming that this heat is expended in melting, the local melt speed is
\begin{equation}
  V_m(t)=\frac{q_w(t)}{\rho_i\Gamma},
  \label{eq:app_stefan}
\end{equation}
where $\rho_i$ is the ice density and $\Gamma$ is the latent heat of fusion. The erosion accumulated over a duration $\tau$ is therefore
\begin{equation}
  X(\tau)=\frac{1}{\rho_i\Gamma}\int_0^\tau q_w(t)\,\mathrm{d}t
  =\frac{\overline{q}_w\tau}{\rho_i\Gamma}.
  \label{eq:app_erosion}
\end{equation}
Consequently, two methods that give the same time-mean heat flux also give the same erosion over the same interval. No additional approximation is introduced in converting $q_w$ to $X$.

\subsection*{A2.\enspace Analytical velocity-spectrum route}

The analytical route asks what single equivalent wave carries the same velocity variance as the full spectrum. For an elevation spectrum $S_\eta(\omega)$ normalised such that
\begin{equation}
  \int_0^\infty S_\eta(\omega)\,\mathrm{d}\omega=\frac{H_s^2}{16},
\end{equation}
the horizontal-velocity spectrum at the waterline is
\begin{equation}
  S_u(\omega)=\left|\mathcal{T}_u(\omega)\right|^2S_\eta(\omega),
  \qquad
  \mathcal{T}_u(\omega)=\omega\frac{\cosh(kd)}{\sinh(kd)},
  \label{eq:app_velocity_spectrum}
\end{equation}
where $k$ satisfies the linear dispersion relation. The r.m.s. velocity and the velocity-spectrum-weighted angular frequency are
\begin{equation}
  U_{\mathrm{rms}}=\left[\int_0^\infty S_u(\omega)\,\mathrm{d}\omega\right]^{1/2},
  \qquad
  \overline{\omega}_u=
  \frac{\int_0^\infty \omega S_u(\omega)\,\mathrm{d}\omega}
       {\int_0^\infty S_u(\omega)\,\mathrm{d}\omega},
  \label{eq:app_equivalent_parameters}
\end{equation}
and the corresponding representative orbital excursion is $a_{\mathrm{eq}}=U_{\mathrm{rms}}/\overline{\omega}_u$. For a zero-mean Gaussian velocity process,
\begin{equation}
  \mathbb{E}[|u|]=\sqrt{\frac{2}{\pi}}\,U_{\mathrm{rms}}.
  \label{eq:app_gaussian_velocity}
\end{equation}

The general Jonsson/Nunner closure is then evaluated at $a_{\mathrm{eq}}$. The rough-wall friction coefficient is obtained from the Lambert-$W$ solution derived in Paper~1,
\begin{equation}
  C_{f,\mathrm{eq}}=
  \left[
    \frac{\ln 10}{4W\!\left(1.916\,a_{\mathrm{eq}}/k_s\right)}
  \right]^2.
  \label{eq:app_cf_lambert}
\end{equation}
Defining
\begin{equation}
  \mathrm{Re}_{0,\mathrm{eq}}=\frac{U_{\mathrm{rms}}a_{\mathrm{eq}}}{\nu},
  \qquad
  C_{f0,\mathrm{eq}}=0.09\,\mathrm{Re}_{0,\mathrm{eq}}^{-0.2},
\end{equation}
the smooth-wall Stanton number and its rough-wall adjustment are
\begin{align}
  \mathrm{St}_{0,\mathrm{eq}}
  &=\frac{0.5C_{f0,\mathrm{eq}}}
  {1+12.8\left(\mathrm{Pr}^{0.68}-1\right)\sqrt{0.5C_{f0,\mathrm{eq}}}}, \\
  \mathrm{St}_{\mathrm{eq}}
  &=\mathrm{St}_{0,\mathrm{eq}}
  \sqrt{\frac{C_{f,\mathrm{eq}}}{C_{f0,\mathrm{eq}}}}.
  \label{eq:app_stanton_equivalent}
\end{align}
The analytical time-mean heat flux and erosion are therefore
\begin{equation}
  \overline{q}_{w,A}=\rho_wc_p\Delta T_{\mathrm{wi}}\,\mathrm{St}_{\mathrm{eq}}
  \sqrt{\frac{2}{\pi}}\,U_{\mathrm{rms}},
  \qquad
  X_A=\frac{\overline{q}_{w,A}\tau}{\rho_i\Gamma}.
  \label{eq:app_route_a}
\end{equation}

\subsection*{A3.\enspace Numerical zero-upcrossing route}

The numerical route makes no such reduction: it generates the sea surface directly and lets individual waves emerge from zero upcrossings. Mutually consistent elevation and velocity histories share the same random phases:
\begin{equation}
  \eta(t)=\sum_{n=1}^{N_f}a_n\cos(\omega_nt+\varphi_n),
  \qquad
  u(t)=\sum_{n=1}^{N_f}\mathcal{T}_u(\omega_n)a_n
  \cos(\omega_nt+\varphi_n),
  \label{eq:app_time_series}
\end{equation}
where $a_n=\sqrt{2S_\eta(\omega_n)\Delta\omega}$. Successive zero upcrossings divide the record into waves of duration $T_j$. For each wave,
\begin{equation}
  \overline{U}_j=\frac{1}{T_j}\int_{t_j}^{t_j+T_j}|u(t)|\,\mathrm{d}t,
  \qquad
  \omega_j=\frac{2\pi}{T_j},
  \qquad
  a_j=\frac{\overline{U}_j}{\omega_j}.
  \label{eq:app_wave_parameters}
\end{equation}
The same Jonsson/Nunner closure used in Route~A is evaluated with $(\overline{U}_j,a_j)$ to obtain $\mathrm{St}_j$ for every wave. Time averaging then gives
\begin{equation}
  \overline{q}_{w,B}=
  \frac{\displaystyle\sum_j
  \rho_wc_p\Delta T_{\mathrm{wi}}\,\mathrm{St}_j\overline{U}_jT_j}
  {\displaystyle\sum_jT_j},
  \qquad
  X_B=\frac{\overline{q}_{w,B}\sum_jT_j}{\rho_i\Gamma}.
  \label{eq:app_route_b}
\end{equation}

The two routes were compared using \texttt{equivalenceTest.m} for a Pierson--Moskowitz spectrum ($\gamma=1$, $H_s=2$~m, $T_p=8$~s, $d=10$~m) represented by a 2000~s random-phase history. For the representative realisation, the final heat fluxes differed by approximately 0.5\%. Equation~\eqref{eq:app_erosion} means that the corresponding erosion estimates differ by the same relative amount. This agreement verifies the zero-upcrossing implementation without invoking an assumed analytical distribution of individual wave heights.

\subsection*{A4.\enspace Compact field-scale implementation}

The field simulations use a compact form of this closure; this subsection records what is actually implemented and why it suffices. They use the power-law approximation
\begin{equation}
  C_f=0.138\left(\frac{a_{1m}}{k_s}\right)^{-0.4},
  \label{eq:app_cf_power}
\end{equation}
which reproduces the Lambert-$W$ solution to within approximately 7\% over $a_{1m}/k_s\approx30$--$300$ \citep{lu2025}. This is the field-scale excursion-to-roughness regime for which the compact model is applied here. The approximation changes the evaluation of the closure, but not the spectral logic or the common conversion from heat flux to erosion.

At the waterline, the resulting non-breaking erosion per wave is proportional to $H_j^{0.8}$ (Eq.~\ref{eq:Xm_Tcancel}). For an interval of duration $\tau$, the directly computed irregular-to-regular ratio is therefore
\begin{equation}
  \mathcal{R}_{\tau}
  =\frac{X_{ir,\mathrm{nb}}}{X_{r,\mathrm{nb}}}
  =\frac{T_p}{\tau}
   \frac{\displaystyle\sum_j H_j^{0.8}}{H_s^{0.8}}.
  \label{eq:app_field_ratio}
\end{equation}
Introducing the nondimensional variables $\widehat{H}_j=H_j/H_s$ and $\widehat{\tau}=\tau/T_p$ gives
\begin{equation}
  \mathcal{R}_{\tau}
  =\frac{1}{\widehat{\tau}}
   \sum_j\widehat{H}_j^{0.8}.
  \label{eq:app_nondim_ratio}
\end{equation}
For the fixed $\gamma=1$ Pierson--Moskowitz spectral shape used throughout this study, the nondimensional zero-upcrossing population is statistically invariant under changes in $H_s$ and $T_p$. The ensemble or long-record limit of Eq.~\eqref{eq:app_nondim_ratio} is therefore a constant: the correction is independent of the dimensional sea-state scale and of $C_{\mathrm{nb}}$, $k_s$, and $\Delta T_{\mathrm{wi}}$. The numerical realisation used here gives $\mathcal{R}\approx0.86$, corresponding to approximately 14\% less erosion in T2 than T1. Individual one-hour realisations fluctuate around this value because of finite sampling. The factor is therefore scale-invariant with respect to $H_s$ and $T_p$ for the adopted Pierson--Moskowitz spectrum and power-law closure; a different spectral shape, peak-enhancement factor, or melt closure can produce a different correction.

\section{S1-guided Sentinel-2 plateau-break detection}
\label{app:s2_detection}

This appendix documents the detection algorithm summarised in Sect.~\ref{sec:detections}. The analysis uses cropped three-band Sentinel-2 image stacks (red B4, green B3, and near-infrared B8), with the processed-scene manifest supplied in the data archive.

The detection uses a bidirectional ``plateau-break'' algorithm (Fig.~\ref{fig:s2_probe}). We seed the manually identified S1 reference coastline every 20\,m. From each seed point, we sample the near-infrared (B8) field at 5\,m intervals along a probing line normal to the coast, extending up to 1~km in both the seaward and landward directions so that the actual S2 coastline may reside on either side of the S1 baseline.

Along each probing line, we extract the near-infrared (B8) digital numbers (DN). In the processed scenes, the NIR band provides the strongest ice--ocean contrast: the ice-shelf surface forms a high-DN plateau that drops abruptly to low open-water values. Walking outward from the ice plateau along the probing line, the algorithm identifies the calving front as the first point where the NIR value drops below 85\% of the estimated plateau level and shows a strongly negative gradient; the exact detection thresholds are documented in the released code.

If the algorithm fails to detect such a clear gradient break (e.g. if the seed point resides over open water following substantial retreat), a fallback search identifies the strongest gradient minimum landward of the seed. Detections deviating by more than 200~m from the local median offset are rejected, and the S1 baseline position is retained at those stations. Finally, we assign a confidence level to each accepted point based on the NIR-gradient magnitude (represented by the colour scheme in Fig.~\ref{fig:s2_probe}), where strong, sharp transitions indicate a clean, unobstructed calving face.

\subsection*{B1.\enspace Comparison against the Sentinel-1 reference}

\textbf{Cross-sensor comparison.}
To evaluate the high-resolution structural details captured by our automated plateau-break algorithm against the manual S1 reference, we measure the geometric distance between the two coastlines. We use the adaptive perpendicular (chainage) measurement technique (detailed in Sect.~\ref{sec:retreat_methods}) to calculate the offset at each point, with results summarised in Table~\ref{tab:s2_validation}. 

\begin{table}[t]
\caption{Comparison of automatically detected Sentinel-2 ice fronts against
Sentinel-1 reference delineations. Signed co-registration offsets (positive = S2 more seaward
than S1) along the 8.9~km section. Median $|\mathrm{grad}|$ is the detection-quality
metric (DN\,m$^{-1}$); values $\geq30$ indicate strong ice--ocean transitions.}
\label{tab:s2_validation}
\begin{tabular}{llrrrr}
\hline
S2 date & S1 reference & Mean (m) & Std (m) & Median (m) & Med $|\mathrm{grad}|$ \\
\hline
7 Jan 2024  & 7 Jan 2024  & $-16$ & 27 & $-5$ & 42 \\
8 Jan 2024  & 7 Jan 2024  & $-16$ & 26 & $-5$ & 43 \\
17 Jan 2024 & 16 Jan 2024 & $-51$ & 66 & $-6$ & 60 \\
27 Jan 2024 & 26 Jan 2024 & $-5$  & 26 & $-4$ & 47 \\
27 Feb 2024 & 28 Feb 2024 & $+18$ & 15 & $+29$ & 29 \\
\hline
\end{tabular}
\end{table}

The objective is not to reproduce the S1 lines exactly. The near-zero median offsets ($\pm 5$\,m for the January scenes) indicate little systematic section-scale displacement between the two products on those dates. The S2 algorithm operates at higher spatial resolution (10\,m versus $\approx$\,40\,m for S1) and applies a repeatable detection rule, but neither product is treated as error-free. Their standard deviations of 26--66\,m combine differences in sensor response, front delineation, acquisition time, and genuine fine-scale front geometry.



\noappendix

\codedataavailability{
ERA5 wave reanalysis data were obtained from the
\href{https://cds.climate.copernicus.eu}{Copernicus Climate Data Store}
(last access: March 2026;
\citealp{hersbach2020era5}).
Sentinel-1 and Sentinel-2 imagery were obtained from the
\href{https://dataspace.copernicus.eu}{Copernicus Data Space Ecosystem}
(last access: March 2026).
NPI manually digitised Sentinel-1 ice-front polylines were provided by the Norwegian
Polar Institute (NPI). The model code, satellite-analysis code, processed front
positions, and derived retreat series are openly available at
\href{https://github.com/KongDAFFY/ice-shelf-wave-erosion-code}{github.com/KongDAFFY/ice-shelf-wave-erosion-code}
and permanently archived on Zenodo (\url{https://doi.org/10.5281/zenodo.21622437}).
}

\authorcontribution{
W.~L.: Conceptualisation, Methodology, Software, Formal analysis, Data curation, Investigation, Writing -- original draft, Writing -- review and editing.
L.~W.: Data curation (manual digitisation of the Sentinel-1 ice-front polylines), Resources, Writing -- review and editing.
B.~G.: Investigation, Resources, Writing -- review and editing.
S.~J.: Formal analysis, Writing -- review and editing.
D.~M.: Conceptualisation, Investigation, Resources, Writing -- review and editing.
M.~F.: Investigation, Writing -- review and editing.
R.~Lambert: Investigation, Writing -- review and editing.
H.~G.: Conceptualisation, Resources, Writing -- review and editing.
G.~M.: Resources, Data curation, Writing -- review and editing.
R.~Lubbad: Supervision, Writing -- review and editing.
S.~L.: Supervision, Writing -- review and editing.
}

\competinginterests{The authors declare that they have no competing interests.}


\begin{acknowledgements}
ERA5 data were downloaded from the Copernicus Climate Data Store.
Sentinel-1 and Sentinel-2 data were provided by ESA through the Copernicus programme.
The authors thank the Norwegian Polar Institute for providing the manually digitised
Sentinel-1 ice-front polylines used as reference in this study.
This research was supported by internal research funding from the Norwegian University of
Science and Technology (NTNU) and the Norwegian Polar Institute (NPI).
The wave-tank experimental campaign was carried out at the M2C Laboratory
(Morphodynamique Continentale et C\^{o}ti\`{e}re, UMR CNRS 6143,
Universit\'e de Caen Normandie); the authors thank CNRS Caen M2C for access to the facility.
During the preparation of this work, the authors used LLMs to support language editing,
manuscript organisation, consistency checks, LaTeX preparation and coding assistance.
The authors reviewed and edited the content as needed and take full responsibility for
the content of the article.
\end{acknowledgements}


\bibliographystyle{copernicus}
\bibliography{paper2_references}

\end{document}